\documentclass[aps,prl,twocolumn,superscriptaddress]{revtex4-1}

\usepackage{graphicx}
\usepackage{CJK}
\usepackage{dcolumn}
\usepackage{bm}
\usepackage{hyperref}
\usepackage{todonotes}
\usepackage{color}
\usepackage{comment}
\usepackage{sidecap}
\usepackage{nicefrac}
\sidecaptionvpos{figure}{c}

\begin{document}

\title{Remarkable Stability of Charge Density Wave Order in  La$_{1.875}$Ba$_{0.125}$CuO$_4$}

\author{X. M. Chen}
\email{xmchen@bnl.gov; Contributed equally to this work}
\author{V. Thampy}
\email{vthampy@bnl.gov; Contributed equally to this work}
\affiliation{Condensed Matter Physics and Materials Science Department, Brookhaven National Laboratory, Upton, New York 11973, USA}

\author{C. Mazzoli}
\author{A. M. Barbour}
\affiliation{National Synchrotron Light Source II, Brookhaven National Laboratory, Upton, New York 11973, USA}

\author{H. Miao}
\author{G. D. Gu}
\author{Y. Cao}
\affiliation{Condensed Matter Physics and Materials Science Department, Brookhaven National Laboratory, Upton, New York 11973, USA}

\author{J. M. Tranquada}
\author{M. P. M. Dean}
\email{mdean@bnl.gov}
\affiliation{Condensed Matter Physics and Materials Science Department, Brookhaven National Laboratory, Upton, New York 11973, USA}
\author{S. B. Wilkins}
\email{swilkins@bnl.gov }
\affiliation{National Synchrotron Light Source II, Brookhaven National Laboratory, Upton, New York 11973, USA}

\def\mathbi#1{\ensuremath{\textbf{\em #1}}}
\def\Q{\ensuremath{\mathbi{Q}}}
\def\LBCO{La$_{2-x}$Ba$_x$CuO$_4$}
\def\LSCO{La$_{2-x}$Sr$_x$CuO$_4$}
\def\LNSCO{La$_{2-x-y}$Nd$_y$Sr$_x$CuO$_4$}

\newcommand{\microns}{$\mathrm{\mu}$m}
\newcommand{\angstrom}{\mbox{\normalfont\AA}}
\date{\today}

\begin{abstract}
The occurrence of charge-density-wave (CDW) order in underdoped cuprates is now well established, although the precise nature of the CDW and its relationship with superconductivity is not.  Theoretical proposals include contrasting ideas such as that pairing may be driven by CDW fluctuations or that static CDWs may intertwine with a spatially-modulated superconducting wave function.  We test the dynamics of CDW order in La$_{1.825}$Ba$_{0.125}$CuO$_4$ by using x-ray photon correlation spectroscopy (XPCS) at the CDW wave vector, detected resonantly at the Cu $L_3$-edge.  We find that the CDW domains are strikingly static, with no evidence of significant fluctuations up to 2\,\nicefrac{3}{4} hours.  We discuss the implications of these results for some of the competing theories.

\end{abstract}


%
\maketitle

A variety of experiments have now established that charge-density-wave (CDW) order coexists with superconductivity in essentially all underdoped cuprates \cite{Tranquada1995,Hoffman2002,Leboeuf2007,Sebastian2008,Wu2011,Ghiringhelli2012,Chang2012,daSilvaNeto2014,Comin2014,fuji14b,Thampy2014,niels14,crof14,Tabis2014,comi16,campi2015inhomogeneity}.  Conventionally, one expects that the CDW order would compete with superconductivity for electronic density of states near the Fermi level, so the exact nature of the CDW and its relationship with superconductivity has come under great scrutiny. A common scenario in trying to understand the cuprates is that quantum fluctuations near a quantum critical point may drive electron pairing and superconductivity \cite{varm97,cast97,sach99d,vojt00b,chak01,sach03b,aban03}, and some have pointed towards CDW  \cite{cast97,vojt00b,wang15b,Caprara2016} or nematic \cite{maie14,lede15,metl15} fluctuations as the relevant modes. An alternative approach proposes that static CDW order and superconductivity may work together, rather than compete, with the potential for a spatially-modulated superconducting wave function \cite{hime02,Berg2007,corb14,lee14,Fradkin2015,zaan01}. Hence, it is of interest to test whether the CDW in the cuprates is static or fluctuating \cite{LeTacon2014,wu15}. To address this question, we have chosen to study La$_{1.825}$Ba$_{0.125}$CuO$_4$ (LBCO~1/8), which has the strongest CDW signal among the cuprates. Even though this optimization of CDW order at $x=1/8$ correlates with a strong depression of the bulk superconducting transition temperature \cite{Fujita2004,Hucker2011}, a two-dimensional (2D) superconducting order parameter amplitude develops at the surprisingly high temperature of 40~K, with phase order becoming established at 16~K \cite{li07}.  Now, if there is only one pairing mechanism, and one type of CDW, in the cuprates then LBCO~1/8, having the strongest CDW signal in diffraction \cite{Abbamonte2005, Wilkins2011,Hucker2013, Thampy2013, DeanLBCO2013}, provides a sensitive test of the nature of the CDW and its relationship with superconductivity.  

In this paper, we use x-ray photon correlation spectroscopy (XPCS) to test for possible slow CDW fluctuations in LBCO~1/8. In XPCS, coherent scattering from the microstructure disorder, or domains, results in a complex interference (``speckle'') pattern. By tracking the evolution of the speckle pattern with time, one can infer the underlying domain dynamics \cite{DaintyBook, Brauer1995, Sutton2008}. This technique has been used to detect mesoscopic fluctuations associated with CDWs in systems such as chromium \cite{Shpyrko2007} and TaS$_2$ \cite{Su2012}.  When fluctuations are faster than the time for a single measurement, the size of the interference contrast is reduced \cite{sutt02, Pinsolle2012}. 

An innovative aspect of our work is that we use coherent soft x-rays tuned to the Cu $L_3$ and O $K$ absorption edges to enhance the intensity of the CDW peak \cite{Abbamonte2005}.  We find that the detected speckle pattern is remarkably stable over timescales from minutes to hours at temperatures up to at least 90\%\ of the disordering transition temperature. Furthermore, the magnitude of the interference contrast is consistent with perfectly coherent scattering from the sample, so that there is no evidence for significant fluctuations of the CDW domains on a timescale faster than the measurement.

\begin{figure}
    \includegraphics[width=0.48\textwidth]{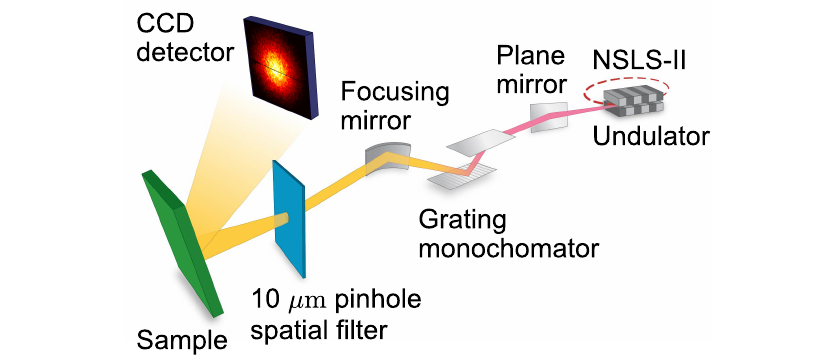}
    \caption{(Color online) Measurement configuration at the 23-ID-1 (CSX-1) beamline at NSLS-II. X-rays from the undulator are energy-dispersed and focused onto a 10~\microns{} pinhole $\sim 5$~mm from the sample. This pinhole acts like an monochromator exit slit achieving $\sim 2$~\microns{} longitudinal coherence and defines the beamsize and transverse coherence length at the sample. Bragg speckle patterns are measured using a charge coupled device (CCD) based device capable of 100 frames per second \cite{fccd_camera}.}
    \label{CSX}
\end{figure}

Figure~\ref{CSX} outlines the experimental setup at the 23-ID-1 beamline at the National Syncrotron Light Source II which is optimized to deliver extremely high coherent flux ($\sim 10^{13}$~photons/s) at the sample.  Following the 10~\microns{} pinhole, the beam has essentially perfect transverse coherence and a longitudinal coherence length of $\sim 2$~\microns{}. Data were collected using a Fast CCD \cite{fccd_camera} with a maximum readout rate of 100~Hz and 30$\times$30~\microns{}$^2$ pixel size placed 340~mm from the sample. Single crystals of LBCO~1/8 were grown using the floating zone method, and characterized extensively in previous studies \cite{Wilkins2011, Thampy2013, DeanLBCO2013, Dean2015}.  Here we report wavevectors in terms of the HTT crystal structure with $a=b=3.78$ and $c=13.28$~\AA{}. In this notation, the CDW occurs at a wavevector of $(\pm 0.236,0,0.5)$ in $(H,K,L)$ reciprocal lattice units, where positive $H$ values correspond to larger incident x-ray angles. The sample was cleaved \emph{ex-situ} and mounted with a $[H,0,L]$ scattering plane.  All data were collected with horizontal ($\sigma$)-polarized incident x-rays.

\begin{figure}
    \includegraphics[width=0.475\textwidth]{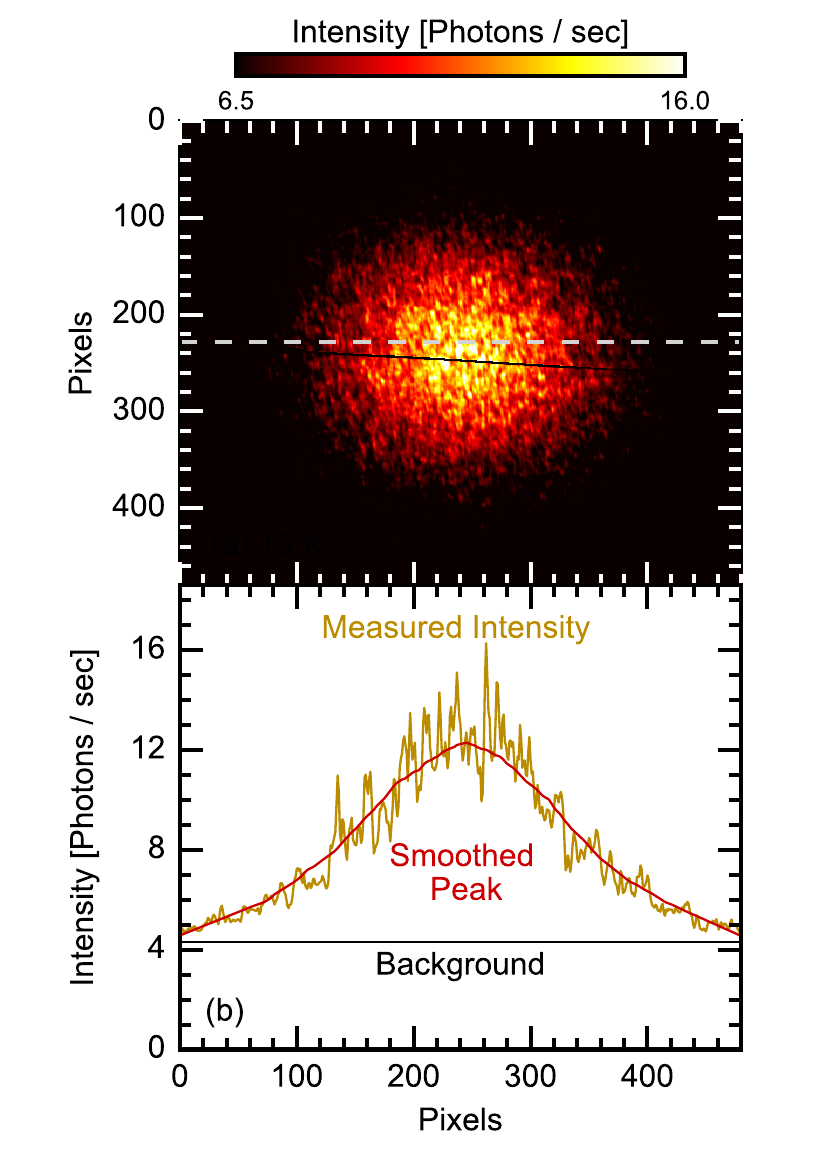}
    \caption{(Color online) (a) A detector image from the LBCO~1/8 CDW Bragg peak at 15~K. The black diagonal line is an artifact from the beamstop. (b) A line cut of the measured intensity taken on the dotted line in panel (a) is shown in yellow. Red and black lines indicate the smoothed peak envelope and the fluorescence-dominated background respectively. The speckle-modulations on top of the peak arise from coherent interference between different CDW domains.}
    \label{Speckle_visibility}
\end{figure}

Figure~\ref{Speckle_visibility}(a) plots the detector image obtained at $(-0.236,0,1.5)$ at 15~K using the Cu $L_3$-edge resonance. Intensity modulations, or speckles, are clearly visible on top of the CDW Bragg peak and serve as a fingerprint of the domain configuration. The line cut in Fig.~\ref{Speckle_visibility}(b) shows the separate components to the scattering. In Bragg coherent scattering measurements, the speckle intensity is modulated by an overall envelope calculated here by applying the Savitzky-Golay smoothing method. The CDW peak sits on top of an approximately flat background, primarily from x-ray fluorescence.

\begin{figure*}[t]
    \centering
    \includegraphics[width=\textwidth]{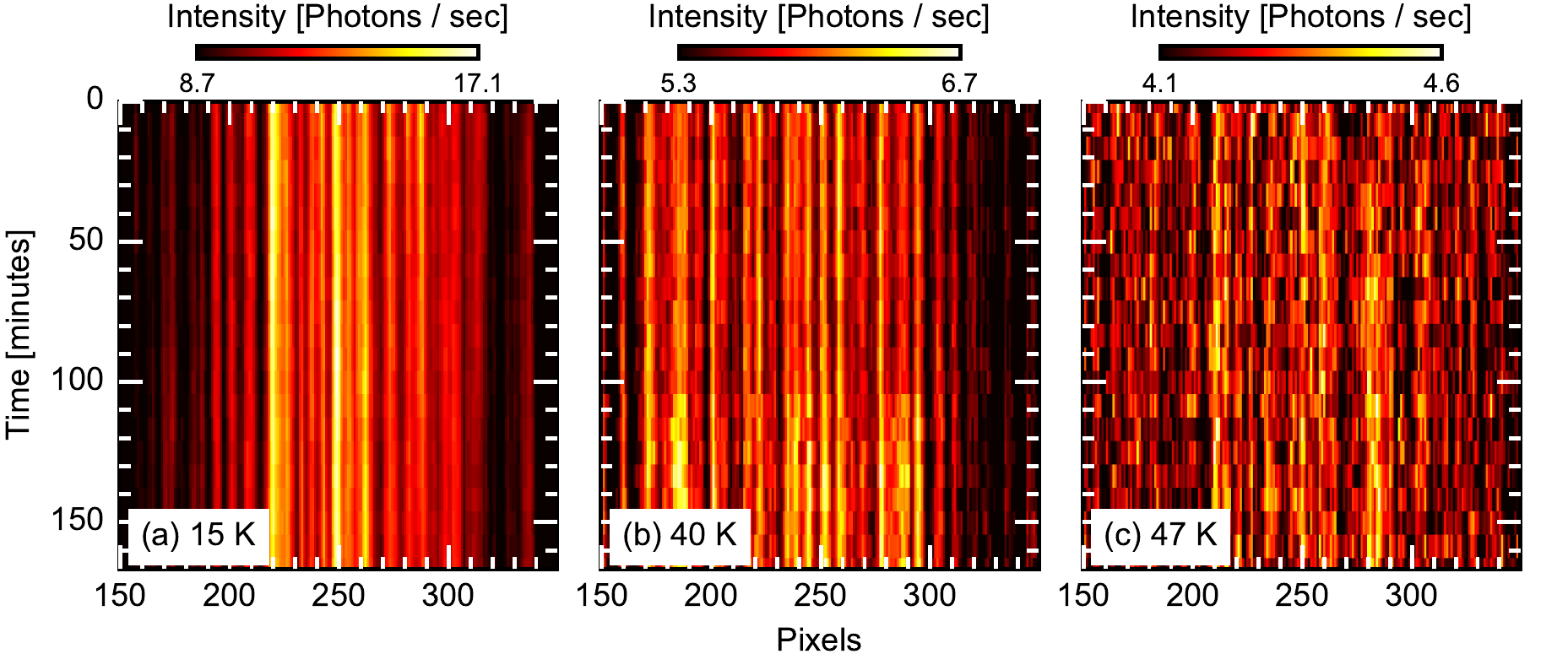}
    \caption{(Color online) Waterfall plots showing horizontal line cuts through the CDW Bragg peak, along the dotted line in Fig.~\ref{Speckle_visibility}(a), as a function of time for (a) 15, (b) 40 and (c) 47~K. The constant vertical streaks in these plots indicate static CDW domains from 15 to 47~K.}
    \label{Waterfalls}
\end{figure*}

The relative strength of the speckles can be quantified by calculating the intensity variation
\begin{equation}
    V = \frac{I_{\text{s}}^{\text{max}} - I_{\text{s}}^{\text{min}}}{I_{\text{s}}^{\text{max}} + I_{\text{s}}^{\text{min}}}
\label{eq:I_speckle}
\end{equation}
where $I_{\text{s}}^{\text{max}}$ and $I_{\text{s}}^{\text{min}}$ are the maximum and minimum speckle-modulated intensities. There is a substantial incoherent background from sample fluorescence that we choose to subtract before evaluating $V$. After averaging over  several cuts close to the dotted horizontal line in 15~K data shown in Fig.~\ref{Speckle_visibility}(a), we find $V=0.15 \pm 0.05$. Similar values are obtained independent of data accumulation time, from 1 minute, the shortest time in which sufficient statistics can be collected, to 2\,\nicefrac{3}{4} hours of total measurement time. The background subtraction allows us to compare the measured $V$ values to the theoretical intensity variation, which was calculated based on the expected x-ray coherence properties, scattering geometry and x-ray penetration depth. The calculation \cite{supplementary} yields $V = 0.15$ consistent with the measured value. Additional measurements at the $+H$ satellite and the O $K$-edge resonance are also consistent with calculated values \cite{supplementary}.

Next we address the time evolution of the observed speckles to test for CDW dynamics, which we examine in ``waterfall'' plots or kymographs shown in Fig.~\ref{Waterfalls}.  These waterfall plots are intensity-time plots, and show the intensity measured along the cut indicated by the dotted line in Fig.~\ref{Speckle_visibility}(a), as a function of time. Each row in these plots has been averaged over 100 images, with an exposure time of 5 seconds per image, adding up to a total of 2\,\nicefrac{3}{4} hours spanned by the vertical time axis. In particular, the 15~K data in Fig.~\ref{Waterfalls}(a) exhibit distinct vertical streaks indicating that the positions of pixels with high intensity, i.e.\ speckles, remain unchanged over the 2\,\nicefrac{3}{4} hour measurement period. This provides the first direct evidence for static CDW domains in LBCO~1/8 over these timescales.

\begin{figure}
    \includegraphics{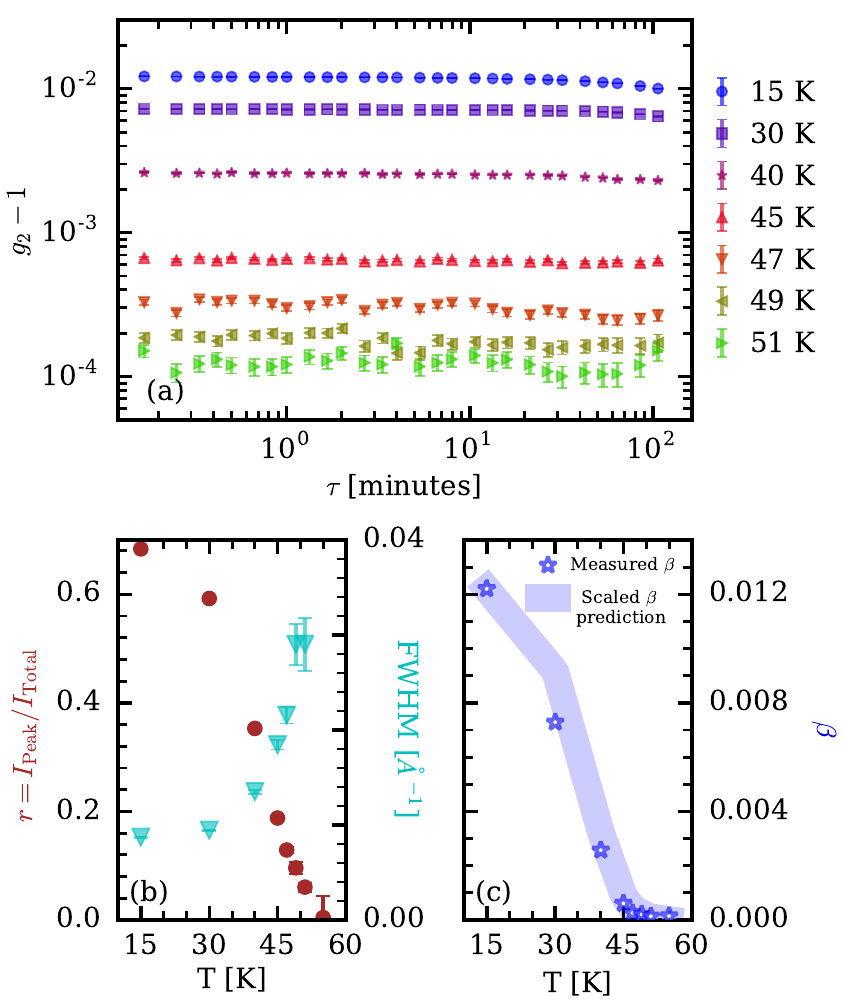}
    \caption{(Color online) (a) Normalized one-time correlation function as a function of time $g_2(\tau)$ from 15-51~K, consistent with static speckles up to 47~K. (b) $r$, the ratio of CDW peak intensity to total intensity including background (brown circles, left axis) and FWHM (cyan inverted triangles, right axis) as a function of temperature obtained from fitting a Lorentzian-squared line shape to line cuts through the peak. (c) The speckle contrast factor, $\beta$ obtained by $g_2(\tau \rightarrow 0)$ (blue stars). The transparent blue line shows the value of $\beta (\mathrm{T})$ obtained by rescaling the $\beta$ value at 15~K by the factor $r^2$, $\beta(\mathrm{T}) = \beta(\mathrm{15~K}) * r(\mathrm{T})^2$, as described in the text. This rescaling accounts for the known drop in the peak intensity relative to the total intensity and captures the observed trend of decreasing $\beta(\mathrm{T})$. Errorbars represent statistical uncertainty and are included whenever the errorbar is larger than the point size.}
    \label{g2}
\end{figure}

Numerous previous studies indicate that the CDW correlation length decreases, in concert with a reduction in peak intensity, as  temperature increases and that the CDW peak eventually disappears around 54~K, close to the low temperature tetragonal (LTT) structural phase transition \cite{Abbamonte2005, Hucker2011, Wilkins2011, DeanLBCO2013, Hucker2013} (see  Fig.~\ref{g2}(b)). Given that fluctuations often play a more important role close to phase transitions, we also measured speckle time-dependence closer to the transition temperature. Figures~\ref{Waterfalls}(b) and (c) show waterfall plots for equivalent measurement times at 40~K and 47~K, respectively. At 40~K, we still see clear vertical streaks spanning the entire 2\,\nicefrac{3}{4} hour measurement duration. At 47~K, although the intensity of the speckles is close to the noise floor, the positions of the brightest speckles still remain unchanged over time, within experimental error.

The speckle time dependence was more formally quantified using the normalized one-time correlation function, $g_2(\mathbi{q}, \tau)$, which auto-correlates pixel intensities with different lag times as
\begin{equation}
    g_2(\mathbi{q}, \tau) = \frac{\langle I(\mathbi{q},t)I(\mathbi{q}, t+\tau)\rangle}{\langle I(\mathbi{q}, t)\rangle ^2} = 1 + \beta \vert F(\mathbi{q},\tau)\vert ^2.
    \label{eq:g2}
\end{equation}
Here, $I(\mathbi{q},t)$ is the \emph{total} intensity in a detector pixel, including background, with a scattering wave-vector $\mathbi{q}$ at time $t$, and $\tau$ is the lag time. $\langle\rangle$ indicates time average as well as ensemble average over all pixels with equivalent \mathbi{q} values for improved statistics. In the data presented here we observed no significant \mathbi{q}-dependence and so we omit \mathbi{q} from further formulae. $F(\tau)$ is the intermediate scattering function encoding the dynamics of the sample \cite{Sutton2008, Shpyrko2014}. The parameter $\beta$, referred to as the speckle contrast factor, is related to the square of the speckle intensity variation $V$ defined above. (If the background is first subtracted from the intensities in Eq.~\ref{eq:g2}, then $\beta=V^2$.) We choose to evaluate $g_2(\tau)$ with the full intensity because unreliable results can occur with background-subtracted intensities when the signal becomes very small and counting statistics are limited.

The value of $\beta$ ranges from 0 (no speckle contrast) to 1 (perfect speckle contrast) depending on the sample behavior, the background strength and the experimental setup. Figure~\ref{g2}(a) plots $g_2(\tau)-1$ calculated on constant intensity contours within 100 pixels on the peak maximum. Time-independent $g_2(\tau)$ behavior over at least one hour is observed all the way up to 47~K. The small drop in $g_2(\tau)$ after approximately 2 hours at 15 and 30~K was deemed to arise from finite experimental stability over these long time periods, based on measurements of the specular reflection from the sample, which provides a static reference. Above 47~K we judge the signal-to-noise to be insufficient for a definitive statement. 

From $g_2$ we can extract the quantity $\beta = g_2(\tau \rightarrow 0) -1$, the temperature dependence of which is plotted in Fig.~\ref{g2}(c).  The strong decrease of $\beta$ with temperature could reflect a reduction in speckle intensity variation $V$ indicating an increase in CDW fluctuations; however, $\beta$ is also sensitive to the ratio, $r(T)$, of peak intensity to total (peak plus background) intensity (plotted in Fig.~\ref{g2}(b)).  Analysis of a simple model shows that $\beta$ should be proportional to $r^2(T)$ \cite{supplementary}.  It follows that, if $V$ is independent of temperature, then $\beta(T)$ is directly proportional to $r^2(T)$. Figure~\ref{g2}(c) shows that $\beta(T)$ indeed scales with $r^2(T)$, indicating that the drop in $\beta(T)$ is due to a reduction in CDW peak intensity. As another check, we computed $V$ at 15, 30 and 40~K, obtaining statistically-consistent values (0.15$\pm$0.05, 0.16$\pm$0.05 and 0.15$\pm$0.05).  Thus, we have no evidence for any increase in fluctuations on approaching the transition. 

This result significantly extends and complements the numerous previous studies that examine charge and spin order and fluctuations in the cuprates. For example, nuclear magnetic resonance (NMR) and nuclear quadrupole (NQR) studies have previously indicated the onset of static order (on a timescale of microseconds) in La$_{1.825}$Ba$_{0.125}$CuO$_4$ below approximately 40~K \cite{goto94, hunt01}. However, the signal in this case is thought to be primarily sensitive to spin (rather than charge) fluctuations \cite{juli01, Pelc2016}. XPCS is one of very few techniques capable for testing for CDW fluctuations on very long timescales on the order of hours, the other being scanning tunnelling spectroscopy (STS) which has observed static CDW signal in a few systems such as Bi$_{2}$Sr$_{2}$Ca$_{n-1}$Cu$_{n}$O$_{2n+4+x}$ \cite{Hoffman2002, Parker2010, Comin2014, daSilvaNeto2015}. However, STS is surface sensitive and has never achieved imaging over a sufficiently large and flat surface area for various non-cleavable cuprates including \LBCO{}\cite{Valla1914}. This is particularly significant in view of work suggesting that the ``214'' family of cuprates might host a qualitatively different type of CDW order \cite{Achkar2016}. Furthermore, XPCS is sensitive to fluctuations that are faster than the measurement time, whereas STS is not. We also note that there is minimal work extending to temperatures close to the CDW transition.

What is the significance of our results?  For comparison, we note that quantum fluctuations were detected by XPCS for long-range-ordered CDW domains in chromium metal \cite{Shpyrko2007}.  A fluctuation time scale of roughly an hour was observed at low temperature, decreasing to the scale of a minute on warming above 100~K despite the relatively high transition temperature of 311~K.  In our case, the correlation length is only $\sim 230$~\AA~ at low temperature.  One might expect that this would be limited by quantum fluctuations, but our measurements imply that static disorder determines the range of correlations.  On warming above 40~K, where the strong 2D superconducting correlations and spin stripe order disappear \cite{Tranquada2008}, the correlation length decreases rapidly; however, the thermal disorder does not show up in any significant way as dynamic fluctuations of the scattering associated with the CDW peak.  

The coexistence of very static CDW order with (2D) superconductivity is hard to reconcile with models in which CDW fluctuations are essential to pairing. Such theories are generally compatible with a small fraction of static CDW order, but our measurements show that the bulk of CDW spectral intensity within our studied wavevector range is static, which does contradict such models.  From the perspective of models in which CDW order only competes with superconductivity, one could ask whether these two phases are present in different parts of the sample.  Measurements of properties such as the thermopower \cite{Tranquada2008} and with techniques such as muon spin rotation \cite{savi05} and nuclear quadrupole resonance \cite{Hunt2001} do not support any significant phase segregation. Theories in which CDW and superconducting orders cooperate are more naturally congruent with our observation of static CDW order. 

It is certainly possible that there are differences in the CDW order in LBCO~1/8 compared to that in other cuprates such as YBa$_2$Cu$_3$O$_{6+x}$ \cite{achk16b}, as the CDW in the former is commensurate with spin-density-wave order \cite{Hucker2011}, while in the latter CDW order develops in the presence of a spin gap \cite{huck14}. It is also possible that fluctuations are more likely at different doping levels or in cuprates such as La$_{2-x}$Sr$_x$CuO$_4$ that have weaker lattice anisotropies \cite{jaco15}. Nevertheless, there is nothing to suggest that the nature of the pairing is not the same in these different compounds, and this condition provides interesting constraints on the pairing mechanism.  Finally, we note that spatially modulated pair density has been observed, commensurate with short-range CDW order, in Bi$_2$Sr$_2$CaCu$_2$O$_{8+\delta}$ \cite{Hamidian2016}, so that perhaps there are strong similarities among the CDW-ordered cuprates after all.

In summary, we describe the first application of XPCS to the charge ordering in the canonical cuprate material LBCO~1/8. Robustly static domains are observed up to our maximum measurement time of 2\,\nicefrac{3}{4} hours from 15 up to at least 47~K, close to the disordering transition of 54~K. We furthermore find that the speckle contrast is consistent with perfectly coherent scattering from the sample, so that there is no evidence for significant fluctuations on a timescale faster than our minimum measurement time of 1 minute. These results show that the CDW correlations statically co-exist with the 2D superconductivity in LBCO~1/8, which lends support to a picture in which these order parameters cooperate rather than compete. Our work establishes XPCS as a new bulk-sensitive probe of CDW domains in the cuprates, with sensitivity to far longer timescales than inelastic scattering \cite{LeTacon2014}. 

\begin{acknowledgments}
We thank Mark Sutton for his insightful comments regarding speckle visibility and Gilberto Fabbris and John Hill for discussion about stripe correlations in the cuprates. Work performed at Brookhaven National Laboratory was supported by the US Department of Energy, Division of Materials Science, under Contract Numbers DE-SC00112704. X-ray scattering measurements by H.M., M.P.M.D.\ and J.M.T.\ were  supported by the Center for Emergent Superconductivity, an Energy Frontier Research Center funded by the U.S. DOE, Office of Basic Energy Sciences. This research used resources at the 23-ID-1 beamline of the National Synchrotron Light Source II, a U.S.\ Department of Energy (DOE) Office of Science User Facility operated for the DOE Office of Science by Brookhaven National Laboratory under Contract No. DE-SC0012704.
\end{acknowledgments}


\begin{thebibliography}{71}%
\makeatletter
\providecommand \@ifxundefined [1]{%
 \@ifx{#1\undefined}
}%
\providecommand \@ifnum [1]{%
 \ifnum #1\expandafter \@firstoftwo
 \else \expandafter \@secondoftwo
 \fi
}%
\providecommand \@ifx [1]{%
 \ifx #1\expandafter \@firstoftwo
 \else \expandafter \@secondoftwo
 \fi
}%
\providecommand \natexlab [1]{#1}%
\providecommand \enquote  [1]{``#1''}%
\providecommand \bibnamefont  [1]{#1}%
\providecommand \bibfnamefont [1]{#1}%
\providecommand \citenamefont [1]{#1}%
\providecommand \href@noop [0]{\@secondoftwo}%
\providecommand \href [0]{\begingroup \@sanitize@url \@href}%
\providecommand \@href[1]{\@@startlink{#1}\@@href}%
\providecommand \@@href[1]{\endgroup#1\@@endlink}%
\providecommand \@sanitize@url [0]{\catcode `\\12\catcode `\$12\catcode
  `\&12\catcode `\#12\catcode `\^12\catcode `\_12\catcode `\%12\relax}%
\providecommand \@@startlink[1]{}%
\providecommand \@@endlink[0]{}%
\providecommand \url  [0]{\begingroup\@sanitize@url \@url }%
\providecommand \@url [1]{\endgroup\@href {#1}{\urlprefix }}%
\providecommand \urlprefix  [0]{URL }%
\providecommand \Eprint [0]{\href }%
\providecommand \doibase [0]{http://dx.doi.org/}%
\providecommand \selectlanguage [0]{\@gobble}%
\providecommand \bibinfo  [0]{\@secondoftwo}%
\providecommand \bibfield  [0]{\@secondoftwo}%
\providecommand \translation [1]{[#1]}%
\providecommand \BibitemOpen [0]{}%
\providecommand \bibitemStop [0]{}%
\providecommand \bibitemNoStop [0]{.\EOS\space}%
\providecommand \EOS [0]{\spacefactor3000\relax}%
\providecommand \BibitemShut  [1]{\csname bibitem#1\endcsname}%
\let\auto@bib@innerbib\@empty
\bibitem [{\citenamefont {{Tranquada}}\ \emph {et~al.}(1995)\citenamefont
  {{Tranquada}}, \citenamefont {{Sternlieb}}, \citenamefont {{Axe}},
  \citenamefont {{Nakamura}},\ and\ \citenamefont {{Uchida}}}]{Tranquada1995}%
  \BibitemOpen
  \bibfield  {author} {\bibinfo {author} {\bibfnamefont {J.~M.}\ \bibnamefont
  {{Tranquada}}}, \bibinfo {author} {\bibfnamefont {B.~J.}\ \bibnamefont
  {{Sternlieb}}}, \bibinfo {author} {\bibfnamefont {J.~D.}\ \bibnamefont
  {{Axe}}}, \bibinfo {author} {\bibfnamefont {Y.}~\bibnamefont {{Nakamura}}}, \
  and\ \bibinfo {author} {\bibfnamefont {S.}~\bibnamefont {{Uchida}}},\ }\href
  {\doibase 10.1038/375561a0} {\bibfield  {journal} {\bibinfo  {journal}
  {Nature}\ }\textbf {\bibinfo {volume} {375}},\ \bibinfo {pages} {561}
  (\bibinfo {year} {1995})}\BibitemShut {NoStop}%
\bibitem [{\citenamefont {Hoffman}\ \emph {et~al.}(2002)\citenamefont
  {Hoffman}, \citenamefont {Hudson}, \citenamefont {Lang}, \citenamefont
  {Madhavan}, \citenamefont {Eisaki}, \citenamefont {Uchida},\ and\
  \citenamefont {Davis}}]{Hoffman2002}%
  \BibitemOpen
  \bibfield  {author} {\bibinfo {author} {\bibfnamefont {J.}~\bibnamefont
  {Hoffman}}, \bibinfo {author} {\bibfnamefont {E.}~\bibnamefont {Hudson}},
  \bibinfo {author} {\bibfnamefont {K.}~\bibnamefont {Lang}}, \bibinfo {author}
  {\bibfnamefont {V.}~\bibnamefont {Madhavan}}, \bibinfo {author}
  {\bibfnamefont {H.}~\bibnamefont {Eisaki}}, \bibinfo {author} {\bibfnamefont
  {S.}~\bibnamefont {Uchida}}, \ and\ \bibinfo {author} {\bibfnamefont
  {J.}~\bibnamefont {Davis}},\ }\href@noop {} {\bibfield  {journal} {\bibinfo
  {journal} {Science}\ }\textbf {\bibinfo {volume} {295}},\ \bibinfo {pages}
  {466} (\bibinfo {year} {2002})}\BibitemShut {NoStop}%
\bibitem [{\citenamefont {LeBoeuf}\ \emph {et~al.}(2007)\citenamefont
  {LeBoeuf}, \citenamefont {Doiron-Leyraud}, \citenamefont {Levallois},
  \citenamefont {Daou}, \citenamefont {Bonnemaison}, \citenamefont {Hussey},
  \citenamefont {Balicas}, \citenamefont {Ramshaw}, \citenamefont {Liang},
  \citenamefont {Bonn} \emph {et~al.}}]{Leboeuf2007}%
  \BibitemOpen
  \bibfield  {author} {\bibinfo {author} {\bibfnamefont {D.}~\bibnamefont
  {LeBoeuf}}, \bibinfo {author} {\bibfnamefont {N.}~\bibnamefont
  {Doiron-Leyraud}}, \bibinfo {author} {\bibfnamefont {J.}~\bibnamefont
  {Levallois}}, \bibinfo {author} {\bibfnamefont {R.}~\bibnamefont {Daou}},
  \bibinfo {author} {\bibfnamefont {J.-B.}\ \bibnamefont {Bonnemaison}},
  \bibinfo {author} {\bibfnamefont {N.}~\bibnamefont {Hussey}}, \bibinfo
  {author} {\bibfnamefont {L.}~\bibnamefont {Balicas}}, \bibinfo {author}
  {\bibfnamefont {B.}~\bibnamefont {Ramshaw}}, \bibinfo {author} {\bibfnamefont
  {R.}~\bibnamefont {Liang}}, \bibinfo {author} {\bibfnamefont
  {D.}~\bibnamefont {Bonn}},  \emph {et~al.},\ }\href@noop {} {\bibfield
  {journal} {\bibinfo  {journal} {Nature}\ }\textbf {\bibinfo {volume} {450}},\
  \bibinfo {pages} {533} (\bibinfo {year} {2007})}\BibitemShut {NoStop}%
\bibitem [{\citenamefont {Sebastian}\ \emph {et~al.}(2008)\citenamefont
  {Sebastian}, \citenamefont {Harrison}, \citenamefont {Palm}, \citenamefont
  {Murphy}, \citenamefont {Mielke}, \citenamefont {Liang}, \citenamefont
  {Bonn}, \citenamefont {Hardy},\ and\ \citenamefont
  {Lonzarich}}]{Sebastian2008}%
  \BibitemOpen
  \bibfield  {author} {\bibinfo {author} {\bibfnamefont {S.~E.}\ \bibnamefont
  {Sebastian}}, \bibinfo {author} {\bibfnamefont {N.}~\bibnamefont {Harrison}},
  \bibinfo {author} {\bibfnamefont {E.}~\bibnamefont {Palm}}, \bibinfo {author}
  {\bibfnamefont {T.}~\bibnamefont {Murphy}}, \bibinfo {author} {\bibfnamefont
  {C.}~\bibnamefont {Mielke}}, \bibinfo {author} {\bibfnamefont
  {R.}~\bibnamefont {Liang}}, \bibinfo {author} {\bibfnamefont
  {D.}~\bibnamefont {Bonn}}, \bibinfo {author} {\bibfnamefont {W.}~\bibnamefont
  {Hardy}}, \ and\ \bibinfo {author} {\bibfnamefont {G.}~\bibnamefont
  {Lonzarich}},\ }\href@noop {} {\bibfield  {journal} {\bibinfo  {journal}
  {Nature}\ }\textbf {\bibinfo {volume} {454}},\ \bibinfo {pages} {200}
  (\bibinfo {year} {2008})}\BibitemShut {NoStop}%
\bibitem [{\citenamefont {{Wu}}\ \emph {et~al.}(2011)\citenamefont {{Wu}},
  \citenamefont {{Mayaffre}}, \citenamefont {{Kr{\"a}mer}}, \citenamefont
  {{Horvati{\'c}}}, \citenamefont {{Berthier}}, \citenamefont {{Hardy}},
  \citenamefont {{Liang}}, \citenamefont {{Bonn}},\ and\ \citenamefont
  {{Julien}}}]{Wu2011}%
  \BibitemOpen
  \bibfield  {author} {\bibinfo {author} {\bibfnamefont {T.}~\bibnamefont
  {{Wu}}}, \bibinfo {author} {\bibfnamefont {H.}~\bibnamefont {{Mayaffre}}},
  \bibinfo {author} {\bibfnamefont {S.}~\bibnamefont {{Kr{\"a}mer}}}, \bibinfo
  {author} {\bibfnamefont {M.}~\bibnamefont {{Horvati{\'c}}}}, \bibinfo
  {author} {\bibfnamefont {C.}~\bibnamefont {{Berthier}}}, \bibinfo {author}
  {\bibfnamefont {W.~N.}\ \bibnamefont {{Hardy}}}, \bibinfo {author}
  {\bibfnamefont {R.}~\bibnamefont {{Liang}}}, \bibinfo {author} {\bibfnamefont
  {D.~A.}\ \bibnamefont {{Bonn}}}, \ and\ \bibinfo {author} {\bibfnamefont
  {M.-H.}\ \bibnamefont {{Julien}}},\ }\href {\doibase 10.1038/nature10345}
  {\bibfield  {journal} {\bibinfo  {journal} {Nature}\ }\textbf {\bibinfo
  {volume} {477}},\ \bibinfo {pages} {191} (\bibinfo {year}
  {2011})}\BibitemShut {NoStop}%
\bibitem [{\citenamefont {Ghiringhelli}\ \emph {et~al.}(2012)\citenamefont
  {Ghiringhelli}, \citenamefont {Le~Tacon}, \citenamefont {Minola},
  \citenamefont {Blanco-Canosa}, \citenamefont {Mazzoli}, \citenamefont
  {Brookes}, \citenamefont {De~Luca}, \citenamefont {Frano}, \citenamefont
  {Hawthorn}, \citenamefont {He}, \citenamefont {Loew}, \citenamefont {Sala},
  \citenamefont {Peets}, \citenamefont {Salluzzo}, \citenamefont {Schierle},
  \citenamefont {Sutarto}, \citenamefont {Sawatzky}, \citenamefont {Weschke},
  \citenamefont {Keimer},\ and\ \citenamefont {Braicovich}}]{Ghiringhelli2012}%
  \BibitemOpen
  \bibfield  {author} {\bibinfo {author} {\bibfnamefont {G.}~\bibnamefont
  {Ghiringhelli}}, \bibinfo {author} {\bibfnamefont {M.}~\bibnamefont
  {Le~Tacon}}, \bibinfo {author} {\bibfnamefont {M.}~\bibnamefont {Minola}},
  \bibinfo {author} {\bibfnamefont {S.}~\bibnamefont {Blanco-Canosa}}, \bibinfo
  {author} {\bibfnamefont {C.}~\bibnamefont {Mazzoli}}, \bibinfo {author}
  {\bibfnamefont {N.~B.}\ \bibnamefont {Brookes}}, \bibinfo {author}
  {\bibfnamefont {G.~M.}\ \bibnamefont {De~Luca}}, \bibinfo {author}
  {\bibfnamefont {A.}~\bibnamefont {Frano}}, \bibinfo {author} {\bibfnamefont
  {D.~G.}\ \bibnamefont {Hawthorn}}, \bibinfo {author} {\bibfnamefont
  {F.}~\bibnamefont {He}}, \bibinfo {author} {\bibfnamefont {T.}~\bibnamefont
  {Loew}}, \bibinfo {author} {\bibfnamefont {M.~M.}\ \bibnamefont {Sala}},
  \bibinfo {author} {\bibfnamefont {D.~C.}\ \bibnamefont {Peets}}, \bibinfo
  {author} {\bibfnamefont {M.}~\bibnamefont {Salluzzo}}, \bibinfo {author}
  {\bibfnamefont {E.}~\bibnamefont {Schierle}}, \bibinfo {author}
  {\bibfnamefont {R.}~\bibnamefont {Sutarto}}, \bibinfo {author} {\bibfnamefont
  {G.~A.}\ \bibnamefont {Sawatzky}}, \bibinfo {author} {\bibfnamefont
  {E.}~\bibnamefont {Weschke}}, \bibinfo {author} {\bibfnamefont
  {B.}~\bibnamefont {Keimer}}, \ and\ \bibinfo {author} {\bibfnamefont
  {L.}~\bibnamefont {Braicovich}},\ }\href {\doibase 10.1126/science.1223532}
  {\bibfield  {journal} {\bibinfo  {journal} {Science}\ }\textbf {\bibinfo
  {volume} {337}},\ \bibinfo {pages} {821} (\bibinfo {year}
  {2012})}\BibitemShut {NoStop}%
\bibitem [{\citenamefont {{Chang}}\ \emph {et~al.}(2012)\citenamefont
  {{Chang}}, \citenamefont {{Blackburn}}, \citenamefont {{Holmes}},
  \citenamefont {{Christensen}}, \citenamefont {{Larsen}}, \citenamefont
  {{Mesot}}, \citenamefont {{Liang}}, \citenamefont {{Bonn}}, \citenamefont
  {{Hardy}}, \citenamefont {{Watenphul}}, \citenamefont {{Zimmermann}},
  \citenamefont {{Forgan}},\ and\ \citenamefont {{Hayden}}}]{Chang2012}%
  \BibitemOpen
  \bibfield  {author} {\bibinfo {author} {\bibfnamefont {J.}~\bibnamefont
  {{Chang}}}, \bibinfo {author} {\bibfnamefont {E.}~\bibnamefont
  {{Blackburn}}}, \bibinfo {author} {\bibfnamefont {A.~T.}\ \bibnamefont
  {{Holmes}}}, \bibinfo {author} {\bibfnamefont {N.~B.}\ \bibnamefont
  {{Christensen}}}, \bibinfo {author} {\bibfnamefont {J.}~\bibnamefont
  {{Larsen}}}, \bibinfo {author} {\bibfnamefont {J.}~\bibnamefont {{Mesot}}},
  \bibinfo {author} {\bibfnamefont {R.}~\bibnamefont {{Liang}}}, \bibinfo
  {author} {\bibfnamefont {D.~A.}\ \bibnamefont {{Bonn}}}, \bibinfo {author}
  {\bibfnamefont {W.~N.}\ \bibnamefont {{Hardy}}}, \bibinfo {author}
  {\bibfnamefont {A.}~\bibnamefont {{Watenphul}}}, \bibinfo {author}
  {\bibfnamefont {M.~V.}\ \bibnamefont {{Zimmermann}}}, \bibinfo {author}
  {\bibfnamefont {E.~M.}\ \bibnamefont {{Forgan}}}, \ and\ \bibinfo {author}
  {\bibfnamefont {S.~M.}\ \bibnamefont {{Hayden}}},\ }\href {\doibase
  10.1038/nphys2456} {\bibfield  {journal} {\bibinfo  {journal} {Nature
  Physics}\ }\textbf {\bibinfo {volume} {8}},\ \bibinfo {pages} {871} (\bibinfo
  {year} {2012})}\BibitemShut {NoStop}%
\bibitem [{\citenamefont {da~Silva~Neto}\ \emph {et~al.}(2014)\citenamefont
  {da~Silva~Neto}, \citenamefont {Aynajian}, \citenamefont {Frano},
  \citenamefont {Comin}, \citenamefont {Schierle}, \citenamefont {Weschke},
  \citenamefont {Gyenis}, \citenamefont {Wen}, \citenamefont {Schneeloch},
  \citenamefont {Xu} \emph {et~al.}}]{daSilvaNeto2014}%
  \BibitemOpen
  \bibfield  {author} {\bibinfo {author} {\bibfnamefont {E.~H.}\ \bibnamefont
  {da~Silva~Neto}}, \bibinfo {author} {\bibfnamefont {P.}~\bibnamefont
  {Aynajian}}, \bibinfo {author} {\bibfnamefont {A.}~\bibnamefont {Frano}},
  \bibinfo {author} {\bibfnamefont {R.}~\bibnamefont {Comin}}, \bibinfo
  {author} {\bibfnamefont {E.}~\bibnamefont {Schierle}}, \bibinfo {author}
  {\bibfnamefont {E.}~\bibnamefont {Weschke}}, \bibinfo {author} {\bibfnamefont
  {A.}~\bibnamefont {Gyenis}}, \bibinfo {author} {\bibfnamefont
  {J.}~\bibnamefont {Wen}}, \bibinfo {author} {\bibfnamefont {J.}~\bibnamefont
  {Schneeloch}}, \bibinfo {author} {\bibfnamefont {Z.}~\bibnamefont {Xu}},
  \emph {et~al.},\ }\href@noop {} {\bibfield  {journal} {\bibinfo  {journal}
  {Science}\ }\textbf {\bibinfo {volume} {343}},\ \bibinfo {pages} {393}
  (\bibinfo {year} {2014})}\BibitemShut {NoStop}%
\bibitem [{\citenamefont {Comin}\ \emph {et~al.}(2014)\citenamefont {Comin},
  \citenamefont {Frano}, \citenamefont {Yee}, \citenamefont {Yoshida},
  \citenamefont {Eisaki}, \citenamefont {Schierle}, \citenamefont {Weschke},
  \citenamefont {Sutarto}, \citenamefont {He}, \citenamefont {Soumyanarayanan},
  \citenamefont {He}, \citenamefont {Le~Tacon}, \citenamefont {Elfimov},
  \citenamefont {Hoffman}, \citenamefont {Sawatzky}, \citenamefont {Keimer},\
  and\ \citenamefont {Damascelli}}]{Comin2014}%
  \BibitemOpen
  \bibfield  {author} {\bibinfo {author} {\bibfnamefont {R.}~\bibnamefont
  {Comin}}, \bibinfo {author} {\bibfnamefont {A.}~\bibnamefont {Frano}},
  \bibinfo {author} {\bibfnamefont {M.~M.}\ \bibnamefont {Yee}}, \bibinfo
  {author} {\bibfnamefont {Y.}~\bibnamefont {Yoshida}}, \bibinfo {author}
  {\bibfnamefont {H.}~\bibnamefont {Eisaki}}, \bibinfo {author} {\bibfnamefont
  {E.}~\bibnamefont {Schierle}}, \bibinfo {author} {\bibfnamefont
  {E.}~\bibnamefont {Weschke}}, \bibinfo {author} {\bibfnamefont
  {R.}~\bibnamefont {Sutarto}}, \bibinfo {author} {\bibfnamefont
  {F.}~\bibnamefont {He}}, \bibinfo {author} {\bibfnamefont {A.}~\bibnamefont
  {Soumyanarayanan}}, \bibinfo {author} {\bibfnamefont {Y.}~\bibnamefont {He}},
  \bibinfo {author} {\bibfnamefont {M.}~\bibnamefont {Le~Tacon}}, \bibinfo
  {author} {\bibfnamefont {I.~S.}\ \bibnamefont {Elfimov}}, \bibinfo {author}
  {\bibfnamefont {J.~E.}\ \bibnamefont {Hoffman}}, \bibinfo {author}
  {\bibfnamefont {G.~A.}\ \bibnamefont {Sawatzky}}, \bibinfo {author}
  {\bibfnamefont {B.}~\bibnamefont {Keimer}}, \ and\ \bibinfo {author}
  {\bibfnamefont {A.}~\bibnamefont {Damascelli}},\ }\href {\doibase
  10.1126/science.1242996} {\bibfield  {journal} {\bibinfo  {journal}
  {Science}\ }\textbf {\bibinfo {volume} {343}},\ \bibinfo {pages} {390}
  (\bibinfo {year} {2014})}\BibitemShut {NoStop}%
\bibitem [{\citenamefont {Fujita}\ \emph {et~al.}(2014)\citenamefont {Fujita},
  \citenamefont {Hamidian}, \citenamefont {Edkins}, \citenamefont {Kim},
  \citenamefont {Kohsaka}, \citenamefont {Azuma}, \citenamefont {Takano},
  \citenamefont {Takagi}, \citenamefont {Eisaki}, \citenamefont {Uchida},
  \citenamefont {Allais}, \citenamefont {Lawler}, \citenamefont {Kim},
  \citenamefont {Sachdev},\ and\ \citenamefont {Davis}}]{fuji14b}%
  \BibitemOpen
  \bibfield  {author} {\bibinfo {author} {\bibfnamefont {K.}~\bibnamefont
  {Fujita}}, \bibinfo {author} {\bibfnamefont {M.~H.}\ \bibnamefont
  {Hamidian}}, \bibinfo {author} {\bibfnamefont {S.~D.}\ \bibnamefont
  {Edkins}}, \bibinfo {author} {\bibfnamefont {C.~K.}\ \bibnamefont {Kim}},
  \bibinfo {author} {\bibfnamefont {Y.}~\bibnamefont {Kohsaka}}, \bibinfo
  {author} {\bibfnamefont {M.}~\bibnamefont {Azuma}}, \bibinfo {author}
  {\bibfnamefont {M.}~\bibnamefont {Takano}}, \bibinfo {author} {\bibfnamefont
  {H.}~\bibnamefont {Takagi}}, \bibinfo {author} {\bibfnamefont
  {H.}~\bibnamefont {Eisaki}}, \bibinfo {author} {\bibfnamefont {S.-i.}\
  \bibnamefont {Uchida}}, \bibinfo {author} {\bibfnamefont {A.}~\bibnamefont
  {Allais}}, \bibinfo {author} {\bibfnamefont {M.~J.}\ \bibnamefont {Lawler}},
  \bibinfo {author} {\bibfnamefont {E.-A.}\ \bibnamefont {Kim}}, \bibinfo
  {author} {\bibfnamefont {S.}~\bibnamefont {Sachdev}}, \ and\ \bibinfo
  {author} {\bibfnamefont {J.~C.~S.}\ \bibnamefont {Davis}},\ }\href {\doibase
  10.1073/pnas.1406297111} {\bibfield  {journal} {\bibinfo  {journal} {Proc.
  Natl. Acad. Sci. USA}\ }\textbf {\bibinfo {volume} {111}},\ \bibinfo {pages}
  {E3026} (\bibinfo {year} {2014})}\BibitemShut {NoStop}%
\bibitem [{\citenamefont {Thampy}\ \emph {et~al.}(2014)\citenamefont {Thampy},
  \citenamefont {Dean}, \citenamefont {Christensen}, \citenamefont {Steinke},
  \citenamefont {Islam}, \citenamefont {Oda}, \citenamefont {Ido},
  \citenamefont {Momono}, \citenamefont {Wilkins},\ and\ \citenamefont
  {Hill}}]{Thampy2014}%
  \BibitemOpen
  \bibfield  {author} {\bibinfo {author} {\bibfnamefont {V.}~\bibnamefont
  {Thampy}}, \bibinfo {author} {\bibfnamefont {M.~P.~M.}\ \bibnamefont {Dean}},
  \bibinfo {author} {\bibfnamefont {N.~B.}\ \bibnamefont {Christensen}},
  \bibinfo {author} {\bibfnamefont {L.}~\bibnamefont {Steinke}}, \bibinfo
  {author} {\bibfnamefont {Z.}~\bibnamefont {Islam}}, \bibinfo {author}
  {\bibfnamefont {M.}~\bibnamefont {Oda}}, \bibinfo {author} {\bibfnamefont
  {M.}~\bibnamefont {Ido}}, \bibinfo {author} {\bibfnamefont {N.}~\bibnamefont
  {Momono}}, \bibinfo {author} {\bibfnamefont {S.~B.}\ \bibnamefont {Wilkins}},
  \ and\ \bibinfo {author} {\bibfnamefont {J.~P.}\ \bibnamefont {Hill}},\
  }\href {\doibase 10.1103/PhysRevB.90.100510} {\bibfield  {journal} {\bibinfo
  {journal} {Phys. Rev. B}\ }\textbf {\bibinfo {volume} {90}},\ \bibinfo
  {pages} {100510} (\bibinfo {year} {2014})}\BibitemShut {NoStop}%
\bibitem [{\citenamefont {{Christensen}}\ \emph {et~al.}(2014)\citenamefont
  {{Christensen}}, \citenamefont {{Chang}}, \citenamefont {{Larsen}},
  \citenamefont {{Fujita}}, \citenamefont {{Oda}}, \citenamefont {{Ido}},
  \citenamefont {{Momono}}, \citenamefont {{Forgan}}, \citenamefont {{Holmes}},
  \citenamefont {{Mesot}}, \citenamefont {{Huecker}},\ and\ \citenamefont
  {{Zimmermann}}}]{niels14}%
  \BibitemOpen
  \bibfield  {author} {\bibinfo {author} {\bibfnamefont {N.~B.}\ \bibnamefont
  {{Christensen}}}, \bibinfo {author} {\bibfnamefont {J.}~\bibnamefont
  {{Chang}}}, \bibinfo {author} {\bibfnamefont {J.}~\bibnamefont {{Larsen}}},
  \bibinfo {author} {\bibfnamefont {M.}~\bibnamefont {{Fujita}}}, \bibinfo
  {author} {\bibfnamefont {M.}~\bibnamefont {{Oda}}}, \bibinfo {author}
  {\bibfnamefont {M.}~\bibnamefont {{Ido}}}, \bibinfo {author} {\bibfnamefont
  {N.}~\bibnamefont {{Momono}}}, \bibinfo {author} {\bibfnamefont {E.~M.}\
  \bibnamefont {{Forgan}}}, \bibinfo {author} {\bibfnamefont {A.~T.}\
  \bibnamefont {{Holmes}}}, \bibinfo {author} {\bibfnamefont {J.}~\bibnamefont
  {{Mesot}}}, \bibinfo {author} {\bibfnamefont {M.}~\bibnamefont {{Huecker}}},
  \ and\ \bibinfo {author} {\bibfnamefont {M.~v.}\ \bibnamefont
  {{Zimmermann}}},\ }\href@noop {} {\bibfield  {journal} {\bibinfo  {journal}
  {ArXiv e-prints}\ } (\bibinfo {year} {2014})},\ \Eprint
  {http://arxiv.org/abs/1404.3192} {arXiv:1404.3192 [cond-mat.supr-con]}
  \BibitemShut {NoStop}%
\bibitem [{\citenamefont {Croft}\ \emph {et~al.}(2014)\citenamefont {Croft},
  \citenamefont {Lester}, \citenamefont {Senn}, \citenamefont {Bombardi},\ and\
  \citenamefont {Hayden}}]{crof14}%
  \BibitemOpen
  \bibfield  {author} {\bibinfo {author} {\bibfnamefont {T.~P.}\ \bibnamefont
  {Croft}}, \bibinfo {author} {\bibfnamefont {C.}~\bibnamefont {Lester}},
  \bibinfo {author} {\bibfnamefont {M.~S.}\ \bibnamefont {Senn}}, \bibinfo
  {author} {\bibfnamefont {A.}~\bibnamefont {Bombardi}}, \ and\ \bibinfo
  {author} {\bibfnamefont {S.~M.}\ \bibnamefont {Hayden}},\ }\href@noop {}
  {\bibfield  {journal} {\bibinfo  {journal} {Phys. Rev. B}\ }\textbf {\bibinfo
  {volume} {89}},\ \bibinfo {pages} {224513} (\bibinfo {year}
  {2014})}\BibitemShut {NoStop}%
\bibitem [{\citenamefont {Tabis}\ \emph {et~al.}(2014)\citenamefont {Tabis},
  \citenamefont {Li}, \citenamefont {Le~Tacon}, \citenamefont {Braicovich},
  \citenamefont {Kreyssig}, \citenamefont {Minola}, \citenamefont {Dellea},
  \citenamefont {Weschke}, \citenamefont {Veit}, \citenamefont {Ramazanoglu}
  \emph {et~al.}}]{Tabis2014}%
  \BibitemOpen
  \bibfield  {author} {\bibinfo {author} {\bibfnamefont {W.}~\bibnamefont
  {Tabis}}, \bibinfo {author} {\bibfnamefont {Y.}~\bibnamefont {Li}}, \bibinfo
  {author} {\bibfnamefont {M.}~\bibnamefont {Le~Tacon}}, \bibinfo {author}
  {\bibfnamefont {L.}~\bibnamefont {Braicovich}}, \bibinfo {author}
  {\bibfnamefont {A.}~\bibnamefont {Kreyssig}}, \bibinfo {author}
  {\bibfnamefont {M.}~\bibnamefont {Minola}}, \bibinfo {author} {\bibfnamefont
  {G.}~\bibnamefont {Dellea}}, \bibinfo {author} {\bibfnamefont
  {E.}~\bibnamefont {Weschke}}, \bibinfo {author} {\bibfnamefont
  {M.}~\bibnamefont {Veit}}, \bibinfo {author} {\bibfnamefont {M.}~\bibnamefont
  {Ramazanoglu}},  \emph {et~al.},\ }\href@noop {} {\bibfield  {journal}
  {\bibinfo  {journal} {Nature communications}\ }\textbf {\bibinfo {volume}
  {5}} (\bibinfo {year} {2014})}\BibitemShut {NoStop}%
\bibitem [{\citenamefont {Comin}\ and\ \citenamefont
  {Damascelli}(2016)}]{comi16}%
  \BibitemOpen
  \bibfield  {author} {\bibinfo {author} {\bibfnamefont {R.}~\bibnamefont
  {Comin}}\ and\ \bibinfo {author} {\bibfnamefont {A.}~\bibnamefont
  {Damascelli}},\ }\href {\doibase 10.1146/annurev-conmatphys-031115-011401}
  {\bibfield  {journal} {\bibinfo  {journal} {Annu. Rev. Condens. Matter
  Phys.}\ }\textbf {\bibinfo {volume} {7}},\ \bibinfo {pages} {369} (\bibinfo
  {year} {2016})}\BibitemShut {NoStop}%
\bibitem [{\citenamefont {Campi}\ \emph {et~al.}(2015)\citenamefont {Campi},
  \citenamefont {Bianconi}, \citenamefont {Poccia}, \citenamefont {Bianconi},
  \citenamefont {Barba}, \citenamefont {Arrighetti}, \citenamefont {Innocenti},
  \citenamefont {Karpinski}, \citenamefont {Zhigadlo}, \citenamefont {Kazakov}
  \emph {et~al.}}]{campi2015inhomogeneity}%
  \BibitemOpen
  \bibfield  {author} {\bibinfo {author} {\bibfnamefont {G.}~\bibnamefont
  {Campi}}, \bibinfo {author} {\bibfnamefont {A.}~\bibnamefont {Bianconi}},
  \bibinfo {author} {\bibfnamefont {N.}~\bibnamefont {Poccia}}, \bibinfo
  {author} {\bibfnamefont {G.}~\bibnamefont {Bianconi}}, \bibinfo {author}
  {\bibfnamefont {L.}~\bibnamefont {Barba}}, \bibinfo {author} {\bibfnamefont
  {G.}~\bibnamefont {Arrighetti}}, \bibinfo {author} {\bibfnamefont
  {D.}~\bibnamefont {Innocenti}}, \bibinfo {author} {\bibfnamefont
  {J.}~\bibnamefont {Karpinski}}, \bibinfo {author} {\bibfnamefont
  {N.}~\bibnamefont {Zhigadlo}}, \bibinfo {author} {\bibfnamefont
  {S.}~\bibnamefont {Kazakov}},  \emph {et~al.},\ }\href@noop {} {\bibfield
  {journal} {\bibinfo  {journal} {Nature}\ }\textbf {\bibinfo {volume} {525}},\
  \bibinfo {pages} {359} (\bibinfo {year} {2015})}\BibitemShut {NoStop}%
\bibitem [{\citenamefont {Varma}(1997)}]{varm97}%
  \BibitemOpen
  \bibfield  {author} {\bibinfo {author} {\bibfnamefont {C.~M.}\ \bibnamefont
  {Varma}},\ }\href {\doibase 10.1103/PhysRevB.55.14554} {\bibfield  {journal}
  {\bibinfo  {journal} {Phys. Rev. B}\ }\textbf {\bibinfo {volume} {55}},\
  \bibinfo {pages} {14554} (\bibinfo {year} {1997})}\BibitemShut {NoStop}%
\bibitem [{\citenamefont {Castellani}\ \emph {et~al.}(1997)\citenamefont
  {Castellani}, \citenamefont {Di~Castro},\ and\ \citenamefont
  {Grilli}}]{cast97}%
  \BibitemOpen
  \bibfield  {author} {\bibinfo {author} {\bibfnamefont {C.}~\bibnamefont
  {Castellani}}, \bibinfo {author} {\bibfnamefont {C.}~\bibnamefont
  {Di~Castro}}, \ and\ \bibinfo {author} {\bibfnamefont {M.}~\bibnamefont
  {Grilli}},\ }\href {\doibase 10.1007/s002570050347} {\bibfield  {journal}
  {\bibinfo  {journal} {Z. Phys. B: Condens. Matter}\ }\textbf {\bibinfo
  {volume} {103}},\ \bibinfo {pages} {137} (\bibinfo {year}
  {1997})}\BibitemShut {NoStop}%
\bibitem [{\citenamefont {Sachdev}(1999)}]{sach99d}%
  \BibitemOpen
  \bibfield  {author} {\bibinfo {author} {\bibfnamefont {S.}~\bibnamefont
  {Sachdev}},\ }\href {\doibase 10.1103/PhysRevB.59.14054} {\bibfield
  {journal} {\bibinfo  {journal} {Phys. Rev. B}\ }\textbf {\bibinfo {volume}
  {59}},\ \bibinfo {pages} {14054} (\bibinfo {year} {1999})}\BibitemShut
  {NoStop}%
\bibitem [{\citenamefont {Vojta}\ \emph {et~al.}(2000)\citenamefont {Vojta},
  \citenamefont {Zhang},\ and\ \citenamefont {Sachdev}}]{vojt00b}%
  \BibitemOpen
  \bibfield  {author} {\bibinfo {author} {\bibfnamefont {M.}~\bibnamefont
  {Vojta}}, \bibinfo {author} {\bibfnamefont {Y.}~\bibnamefont {Zhang}}, \ and\
  \bibinfo {author} {\bibfnamefont {S.}~\bibnamefont {Sachdev}},\ }\href
  {\doibase 10.1103/PhysRevB.62.6721} {\bibfield  {journal} {\bibinfo
  {journal} {Phys. Rev. B}\ }\textbf {\bibinfo {volume} {62}},\ \bibinfo
  {pages} {6721} (\bibinfo {year} {2000})}\BibitemShut {NoStop}%
\bibitem [{\citenamefont {Chakravarty}\ \emph {et~al.}(2001)\citenamefont
  {Chakravarty}, \citenamefont {Laughlin}, \citenamefont {Morr},\ and\
  \citenamefont {Nayak}}]{chak01}%
  \BibitemOpen
  \bibfield  {author} {\bibinfo {author} {\bibfnamefont {S.}~\bibnamefont
  {Chakravarty}}, \bibinfo {author} {\bibfnamefont {R.~B.}\ \bibnamefont
  {Laughlin}}, \bibinfo {author} {\bibfnamefont {D.~K.}\ \bibnamefont {Morr}},
  \ and\ \bibinfo {author} {\bibfnamefont {C.}~\bibnamefont {Nayak}},\ }\href
  {\doibase 10.1103/PhysRevB.63.094503} {\bibfield  {journal} {\bibinfo
  {journal} {Phys. Rev. B}\ }\textbf {\bibinfo {volume} {63}},\ \bibinfo
  {pages} {094503} (\bibinfo {year} {2001})}\BibitemShut {NoStop}%
\bibitem [{\citenamefont {Sachdev}(2003)}]{sach03b}%
  \BibitemOpen
  \bibfield  {author} {\bibinfo {author} {\bibfnamefont {S.}~\bibnamefont
  {Sachdev}},\ }\href {\doibase 10.1103/RevModPhys.75.913} {\bibfield
  {journal} {\bibinfo  {journal} {Rev. Mod. Phys.}\ }\textbf {\bibinfo {volume}
  {75}},\ \bibinfo {pages} {913} (\bibinfo {year} {2003})}\BibitemShut
  {NoStop}%
\bibitem [{\citenamefont {Abanov}\ \emph {et~al.}(2003)\citenamefont {Abanov},
  \citenamefont {Chubukov},\ and\ \citenamefont {Schmalian}}]{aban03}%
  \BibitemOpen
  \bibfield  {author} {\bibinfo {author} {\bibfnamefont {A.}~\bibnamefont
  {Abanov}}, \bibinfo {author} {\bibfnamefont {A.~V.}\ \bibnamefont
  {Chubukov}}, \ and\ \bibinfo {author} {\bibfnamefont {J.}~\bibnamefont
  {Schmalian}},\ }\href {\doibase 10.1080/0001873021000057123} {\bibfield
  {journal} {\bibinfo  {journal} {Adv. Phys.}\ }\textbf {\bibinfo {volume}
  {52}},\ \bibinfo {pages} {119} (\bibinfo {year} {2003})}\BibitemShut
  {NoStop}%
\bibitem [{\citenamefont {Wang}\ and\ \citenamefont
  {Chubukov}(2015)}]{wang15b}%
  \BibitemOpen
  \bibfield  {author} {\bibinfo {author} {\bibfnamefont {Y.}~\bibnamefont
  {Wang}}\ and\ \bibinfo {author} {\bibfnamefont {A.~V.}\ \bibnamefont
  {Chubukov}},\ }\href {\doibase 10.1103/PhysRevB.92.125108} {\bibfield
  {journal} {\bibinfo  {journal} {Phys. Rev. B}\ }\textbf {\bibinfo {volume}
  {92}},\ \bibinfo {pages} {125108} (\bibinfo {year} {2015})}\BibitemShut
  {NoStop}%
\bibitem [{\citenamefont {{Caprara}}\ \emph {et~al.}(2016)\citenamefont
  {{Caprara}}, \citenamefont {{Di Castro}}, \citenamefont {{Seibold}},\ and\
  \citenamefont {{Grilli}}}]{Caprara2016}%
  \BibitemOpen
  \bibfield  {author} {\bibinfo {author} {\bibfnamefont {S.}~\bibnamefont
  {{Caprara}}}, \bibinfo {author} {\bibfnamefont {C.}~\bibnamefont {{Di
  Castro}}}, \bibinfo {author} {\bibfnamefont {G.}~\bibnamefont {{Seibold}}}, \
  and\ \bibinfo {author} {\bibfnamefont {M.}~\bibnamefont {{Grilli}}},\
  }\href@noop {} {\bibfield  {journal} {\bibinfo  {journal} {ArXiv e-prints}\ }
  (\bibinfo {year} {2016})},\ \Eprint {http://arxiv.org/abs/1604.07852}
  {arXiv:1604.07852 [cond-mat.supr-con]} \BibitemShut {NoStop}%
\bibitem [{\citenamefont {Maier}\ and\ \citenamefont
  {Scalapino}(2014)}]{maie14}%
  \BibitemOpen
  \bibfield  {author} {\bibinfo {author} {\bibfnamefont {T.~A.}\ \bibnamefont
  {Maier}}\ and\ \bibinfo {author} {\bibfnamefont {D.~J.}\ \bibnamefont
  {Scalapino}},\ }\href {\doibase 10.1103/PhysRevB.90.174510} {\bibfield
  {journal} {\bibinfo  {journal} {Phys. Rev. B}\ }\textbf {\bibinfo {volume}
  {90}},\ \bibinfo {pages} {174510} (\bibinfo {year} {2014})}\BibitemShut
  {NoStop}%
\bibitem [{\citenamefont {Lederer}\ \emph {et~al.}(2015)\citenamefont
  {Lederer}, \citenamefont {Schattner}, \citenamefont {Berg},\ and\
  \citenamefont {Kivelson}}]{lede15}%
  \BibitemOpen
  \bibfield  {author} {\bibinfo {author} {\bibfnamefont {S.}~\bibnamefont
  {Lederer}}, \bibinfo {author} {\bibfnamefont {Y.}~\bibnamefont {Schattner}},
  \bibinfo {author} {\bibfnamefont {E.}~\bibnamefont {Berg}}, \ and\ \bibinfo
  {author} {\bibfnamefont {S.~A.}\ \bibnamefont {Kivelson}},\ }\href {\doibase
  10.1103/PhysRevLett.114.097001} {\bibfield  {journal} {\bibinfo  {journal}
  {Phys. Rev. Lett.}\ }\textbf {\bibinfo {volume} {114}},\ \bibinfo {pages}
  {097001} (\bibinfo {year} {2015})}\BibitemShut {NoStop}%
\bibitem [{\citenamefont {Metlitski}\ \emph {et~al.}(2015)\citenamefont
  {Metlitski}, \citenamefont {Mross}, \citenamefont {Sachdev},\ and\
  \citenamefont {Senthil}}]{metl15}%
  \BibitemOpen
  \bibfield  {author} {\bibinfo {author} {\bibfnamefont {M.~A.}\ \bibnamefont
  {Metlitski}}, \bibinfo {author} {\bibfnamefont {D.~F.}\ \bibnamefont
  {Mross}}, \bibinfo {author} {\bibfnamefont {S.}~\bibnamefont {Sachdev}}, \
  and\ \bibinfo {author} {\bibfnamefont {T.}~\bibnamefont {Senthil}},\ }\href
  {\doibase 10.1103/PhysRevB.91.115111} {\bibfield  {journal} {\bibinfo
  {journal} {Phys. Rev. B}\ }\textbf {\bibinfo {volume} {91}},\ \bibinfo
  {pages} {115111} (\bibinfo {year} {2015})}\BibitemShut {NoStop}%
\bibitem [{\citenamefont {Himeda}\ \emph {et~al.}(2002)\citenamefont {Himeda},
  \citenamefont {Kato},\ and\ \citenamefont {Ogata}}]{hime02}%
  \BibitemOpen
  \bibfield  {author} {\bibinfo {author} {\bibfnamefont {A.}~\bibnamefont
  {Himeda}}, \bibinfo {author} {\bibfnamefont {T.}~\bibnamefont {Kato}}, \ and\
  \bibinfo {author} {\bibfnamefont {M.}~\bibnamefont {Ogata}},\ }\href@noop {}
  {\bibfield  {journal} {\bibinfo  {journal} {Phys. Rev. Lett.}\ }\textbf
  {\bibinfo {volume} {88}},\ \bibinfo {pages} {117001} (\bibinfo {year}
  {2002})}\BibitemShut {NoStop}%
\bibitem [{\citenamefont {Berg}\ \emph {et~al.}(2007)\citenamefont {Berg},
  \citenamefont {Fradkin}, \citenamefont {Kim}, \citenamefont {Kivelson},
  \citenamefont {Oganesyan}, \citenamefont {Tranquada},\ and\ \citenamefont
  {Zhang}}]{Berg2007}%
  \BibitemOpen
  \bibfield  {author} {\bibinfo {author} {\bibfnamefont {E.}~\bibnamefont
  {Berg}}, \bibinfo {author} {\bibfnamefont {E.}~\bibnamefont {Fradkin}},
  \bibinfo {author} {\bibfnamefont {E.-A.}\ \bibnamefont {Kim}}, \bibinfo
  {author} {\bibfnamefont {S.~A.}\ \bibnamefont {Kivelson}}, \bibinfo {author}
  {\bibfnamefont {V.}~\bibnamefont {Oganesyan}}, \bibinfo {author}
  {\bibfnamefont {J.~M.}\ \bibnamefont {Tranquada}}, \ and\ \bibinfo {author}
  {\bibfnamefont {S.~C.}\ \bibnamefont {Zhang}},\ }\href {\doibase
  10.1103/PhysRevLett.99.127003} {\bibfield  {journal} {\bibinfo  {journal}
  {Phys. Rev. Lett.}\ }\textbf {\bibinfo {volume} {99}},\ \bibinfo {pages}
  {127003} (\bibinfo {year} {2007})}\BibitemShut {NoStop}%
\bibitem [{\citenamefont {Corboz}\ \emph {et~al.}(2014)\citenamefont {Corboz},
  \citenamefont {Rice},\ and\ \citenamefont {Troyer}}]{corb14}%
  \BibitemOpen
  \bibfield  {author} {\bibinfo {author} {\bibfnamefont {P.}~\bibnamefont
  {Corboz}}, \bibinfo {author} {\bibfnamefont {T.~M.}\ \bibnamefont {Rice}}, \
  and\ \bibinfo {author} {\bibfnamefont {M.}~\bibnamefont {Troyer}},\
  }\href@noop {} {\bibfield  {journal} {\bibinfo  {journal} {Phys. Rev. Lett.}\
  }\textbf {\bibinfo {volume} {113}},\ \bibinfo {pages} {046402} (\bibinfo
  {year} {2014})}\BibitemShut {NoStop}%
\bibitem [{\citenamefont {Lee}(2014)}]{lee14}%
  \BibitemOpen
  \bibfield  {author} {\bibinfo {author} {\bibfnamefont {P.~A.}\ \bibnamefont
  {Lee}},\ }\href@noop {} {\bibfield  {journal} {\bibinfo  {journal} {Phys.
  Rev. X}\ }\textbf {\bibinfo {volume} {4}},\ \bibinfo {pages} {031017}
  (\bibinfo {year} {2014})}\BibitemShut {NoStop}%
\bibitem [{\citenamefont {Fradkin}\ \emph {et~al.}(2015)\citenamefont
  {Fradkin}, \citenamefont {Kivelson},\ and\ \citenamefont
  {Tranquada}}]{Fradkin2015}%
  \BibitemOpen
  \bibfield  {author} {\bibinfo {author} {\bibfnamefont {E.}~\bibnamefont
  {Fradkin}}, \bibinfo {author} {\bibfnamefont {S.~A.}\ \bibnamefont
  {Kivelson}}, \ and\ \bibinfo {author} {\bibfnamefont {J.~M.}\ \bibnamefont
  {Tranquada}},\ }\href {\doibase 10.1103/RevModPhys.87.457} {\bibfield
  {journal} {\bibinfo  {journal} {Rev. Mod. Phys.}\ }\textbf {\bibinfo {volume}
  {87}},\ \bibinfo {pages} {457} (\bibinfo {year} {2015})}\BibitemShut
  {NoStop}%
\bibitem [{\citenamefont {Zaanen}\ \emph {et~al.}(2001)\citenamefont {Zaanen},
  \citenamefont {Osman}, \citenamefont {Kruis}, \citenamefont {Nussinov},\ and\
  \citenamefont {Tworzyd{\l}o}}]{zaan01}%
  \BibitemOpen
  \bibfield  {author} {\bibinfo {author} {\bibfnamefont {J.}~\bibnamefont
  {Zaanen}}, \bibinfo {author} {\bibfnamefont {O.~Y.}\ \bibnamefont {Osman}},
  \bibinfo {author} {\bibfnamefont {H.~V.}\ \bibnamefont {Kruis}}, \bibinfo
  {author} {\bibfnamefont {Z.}~\bibnamefont {Nussinov}}, \ and\ \bibinfo
  {author} {\bibfnamefont {J.}~\bibnamefont {Tworzyd{\l}o}},\ }\href@noop {}
  {\bibfield  {journal} {\bibinfo  {journal} {Phil. Mag. B}\ }\textbf {\bibinfo
  {volume} {81}},\ \bibinfo {pages} {1485} (\bibinfo {year}
  {2001})}\BibitemShut {NoStop}%
\bibitem [{\citenamefont {Le~Tacon}\ \emph {et~al.}(2014)\citenamefont
  {Le~Tacon}, \citenamefont {Bosak}, \citenamefont {Souliou}, \citenamefont
  {Dellea}, \citenamefont {Loew}, \citenamefont {Heid}, \citenamefont {Bohnen},
  \citenamefont {Ghiringhelli}, \citenamefont {Krisch},\ and\ \citenamefont
  {Keimer}}]{LeTacon2014}%
  \BibitemOpen
  \bibfield  {author} {\bibinfo {author} {\bibfnamefont {M.}~\bibnamefont
  {Le~Tacon}}, \bibinfo {author} {\bibfnamefont {A.}~\bibnamefont {Bosak}},
  \bibinfo {author} {\bibfnamefont {S.}~\bibnamefont {Souliou}}, \bibinfo
  {author} {\bibfnamefont {G.}~\bibnamefont {Dellea}}, \bibinfo {author}
  {\bibfnamefont {T.}~\bibnamefont {Loew}}, \bibinfo {author} {\bibfnamefont
  {R.}~\bibnamefont {Heid}}, \bibinfo {author} {\bibfnamefont {K.}~\bibnamefont
  {Bohnen}}, \bibinfo {author} {\bibfnamefont {G.}~\bibnamefont
  {Ghiringhelli}}, \bibinfo {author} {\bibfnamefont {M.}~\bibnamefont
  {Krisch}}, \ and\ \bibinfo {author} {\bibfnamefont {B.}~\bibnamefont
  {Keimer}},\ }\href@noop {} {\bibfield  {journal} {\bibinfo  {journal} {Nature
  Physics}\ }\textbf {\bibinfo {volume} {10}},\ \bibinfo {pages} {52} (\bibinfo
  {year} {2014})}\BibitemShut {NoStop}%
\bibitem [{\citenamefont {Wu}\ \emph {et~al.}(2015)\citenamefont {Wu},
  \citenamefont {Mayaffre}, \citenamefont {Kr\"amer}, \citenamefont
  {Horvati\'c}, \citenamefont {Berthier}, \citenamefont {Hardy}, \citenamefont
  {Liang}, \citenamefont {Bonn},\ and\ \citenamefont {Julien}}]{wu15}%
  \BibitemOpen
  \bibfield  {author} {\bibinfo {author} {\bibfnamefont {T.}~\bibnamefont
  {Wu}}, \bibinfo {author} {\bibfnamefont {H.}~\bibnamefont {Mayaffre}},
  \bibinfo {author} {\bibfnamefont {S.}~\bibnamefont {Kr\"amer}}, \bibinfo
  {author} {\bibfnamefont {M.}~\bibnamefont {Horvati\'c}}, \bibinfo {author}
  {\bibfnamefont {C.}~\bibnamefont {Berthier}}, \bibinfo {author}
  {\bibfnamefont {W.~N.}\ \bibnamefont {Hardy}}, \bibinfo {author}
  {\bibfnamefont {R.}~\bibnamefont {Liang}}, \bibinfo {author} {\bibfnamefont
  {D.~A.}\ \bibnamefont {Bonn}}, \ and\ \bibinfo {author} {\bibfnamefont
  {M.-H.}\ \bibnamefont {Julien}},\ }\href
  {http://dx.doi.org/10.1038/ncomms7438} {\bibfield  {journal} {\bibinfo
  {journal} {Nat. Commun.}\ }\textbf {\bibinfo {volume} {6}},\ \bibinfo {pages}
  {6438} (\bibinfo {year} {2015})}\BibitemShut {NoStop}%
\bibitem [{\citenamefont {Fujita}\ \emph {et~al.}(2004)\citenamefont {Fujita},
  \citenamefont {Goka}, \citenamefont {Yamada}, \citenamefont {Tranquada},\
  and\ \citenamefont {Regnault}}]{Fujita2004}%
  \BibitemOpen
  \bibfield  {author} {\bibinfo {author} {\bibfnamefont {M.}~\bibnamefont
  {Fujita}}, \bibinfo {author} {\bibfnamefont {H.}~\bibnamefont {Goka}},
  \bibinfo {author} {\bibfnamefont {K.}~\bibnamefont {Yamada}}, \bibinfo
  {author} {\bibfnamefont {J.~M.}\ \bibnamefont {Tranquada}}, \ and\ \bibinfo
  {author} {\bibfnamefont {L.~P.}\ \bibnamefont {Regnault}},\ }\href {\doibase
  10.1103/PhysRevB.70.104517} {\bibfield  {journal} {\bibinfo  {journal} {Phys.
  Rev. B}\ }\textbf {\bibinfo {volume} {70}},\ \bibinfo {pages} {104517}
  (\bibinfo {year} {2004})}\BibitemShut {NoStop}%
\bibitem [{\citenamefont {H\"ucker}\ \emph {et~al.}(2011)\citenamefont
  {H\"ucker}, \citenamefont {v.~Zimmermann}, \citenamefont {Gu}, \citenamefont
  {Xu}, \citenamefont {Wen}, \citenamefont {Xu}, \citenamefont {Kang},
  \citenamefont {Zheludev},\ and\ \citenamefont {Tranquada}}]{Hucker2011}%
  \BibitemOpen
  \bibfield  {author} {\bibinfo {author} {\bibfnamefont {M.}~\bibnamefont
  {H\"ucker}}, \bibinfo {author} {\bibfnamefont {M.}~\bibnamefont
  {v.~Zimmermann}}, \bibinfo {author} {\bibfnamefont {G.~D.}\ \bibnamefont
  {Gu}}, \bibinfo {author} {\bibfnamefont {Z.~J.}\ \bibnamefont {Xu}}, \bibinfo
  {author} {\bibfnamefont {J.~S.}\ \bibnamefont {Wen}}, \bibinfo {author}
  {\bibfnamefont {G.}~\bibnamefont {Xu}}, \bibinfo {author} {\bibfnamefont
  {H.~J.}\ \bibnamefont {Kang}}, \bibinfo {author} {\bibfnamefont
  {A.}~\bibnamefont {Zheludev}}, \ and\ \bibinfo {author} {\bibfnamefont
  {J.~M.}\ \bibnamefont {Tranquada}},\ }\href {\doibase
  10.1103/PhysRevB.83.104506} {\bibfield  {journal} {\bibinfo  {journal} {Phys.
  Rev. B}\ }\textbf {\bibinfo {volume} {83}},\ \bibinfo {pages} {104506}
  (\bibinfo {year} {2011})}\BibitemShut {NoStop}%
\bibitem [{\citenamefont {Li}\ \emph {et~al.}(2007)\citenamefont {Li},
  \citenamefont {{H\"ucker}}, \citenamefont {Gu}, \citenamefont {Tsvelik},\
  and\ \citenamefont {Tranquada}}]{li07}%
  \BibitemOpen
  \bibfield  {author} {\bibinfo {author} {\bibfnamefont {Q.}~\bibnamefont
  {Li}}, \bibinfo {author} {\bibfnamefont {M.}~\bibnamefont {{H\"ucker}}},
  \bibinfo {author} {\bibfnamefont {G.~D.}\ \bibnamefont {Gu}}, \bibinfo
  {author} {\bibfnamefont {A.~M.}\ \bibnamefont {Tsvelik}}, \ and\ \bibinfo
  {author} {\bibfnamefont {J.~M.}\ \bibnamefont {Tranquada}},\ }\href@noop {}
  {\bibfield  {journal} {\bibinfo  {journal} {Phys. Rev. Lett.}\ }\textbf
  {\bibinfo {volume} {99}},\ \bibinfo {eid} {067001} (\bibinfo {year}
  {2007})}\BibitemShut {NoStop}%
\bibitem [{\citenamefont {{Abbamonte}}\ \emph {et~al.}(2005)\citenamefont
  {{Abbamonte}}, \citenamefont {{Rusydi}}, \citenamefont {{Smadici}},
  \citenamefont {{Gu}}, \citenamefont {{Sawatzky}},\ and\ \citenamefont
  {{Feng}}}]{Abbamonte2005}%
  \BibitemOpen
  \bibfield  {author} {\bibinfo {author} {\bibfnamefont {P.}~\bibnamefont
  {{Abbamonte}}}, \bibinfo {author} {\bibfnamefont {A.}~\bibnamefont
  {{Rusydi}}}, \bibinfo {author} {\bibfnamefont {S.}~\bibnamefont {{Smadici}}},
  \bibinfo {author} {\bibfnamefont {G.~D.}\ \bibnamefont {{Gu}}}, \bibinfo
  {author} {\bibfnamefont {G.~A.}\ \bibnamefont {{Sawatzky}}}, \ and\ \bibinfo
  {author} {\bibfnamefont {D.~L.}\ \bibnamefont {{Feng}}},\ }\href {\doibase
  10.1038/nphys178} {\bibfield  {journal} {\bibinfo  {journal} {Nature
  Physics}\ }\textbf {\bibinfo {volume} {1}},\ \bibinfo {pages} {155} (\bibinfo
  {year} {2005})}\BibitemShut {NoStop}%
\bibitem [{\citenamefont {Wilkins}\ \emph {et~al.}(2011)\citenamefont
  {Wilkins}, \citenamefont {Dean}, \citenamefont {Fink}, \citenamefont
  {H\"ucker}, \citenamefont {Geck}, \citenamefont {Soltwisch}, \citenamefont
  {Schierle}, \citenamefont {Weschke}, \citenamefont {Gu}, \citenamefont
  {Uchida}, \citenamefont {Ichikawa}, \citenamefont {Tranquada},\ and\
  \citenamefont {Hill}}]{Wilkins2011}%
  \BibitemOpen
  \bibfield  {author} {\bibinfo {author} {\bibfnamefont {S.~B.}\ \bibnamefont
  {Wilkins}}, \bibinfo {author} {\bibfnamefont {M.~P.~M.}\ \bibnamefont
  {Dean}}, \bibinfo {author} {\bibfnamefont {J.}~\bibnamefont {Fink}}, \bibinfo
  {author} {\bibfnamefont {M.}~\bibnamefont {H\"ucker}}, \bibinfo {author}
  {\bibfnamefont {J.}~\bibnamefont {Geck}}, \bibinfo {author} {\bibfnamefont
  {V.}~\bibnamefont {Soltwisch}}, \bibinfo {author} {\bibfnamefont
  {E.}~\bibnamefont {Schierle}}, \bibinfo {author} {\bibfnamefont
  {E.}~\bibnamefont {Weschke}}, \bibinfo {author} {\bibfnamefont
  {G.}~\bibnamefont {Gu}}, \bibinfo {author} {\bibfnamefont {S.}~\bibnamefont
  {Uchida}}, \bibinfo {author} {\bibfnamefont {N.}~\bibnamefont {Ichikawa}},
  \bibinfo {author} {\bibfnamefont {J.~M.}\ \bibnamefont {Tranquada}}, \ and\
  \bibinfo {author} {\bibfnamefont {J.~P.}\ \bibnamefont {Hill}},\ }\href
  {\doibase 10.1103/PhysRevB.84.195101} {\bibfield  {journal} {\bibinfo
  {journal} {Phys. Rev. B}\ }\textbf {\bibinfo {volume} {84}},\ \bibinfo
  {pages} {195101} (\bibinfo {year} {2011})}\BibitemShut {NoStop}%
\bibitem [{\citenamefont {H\"ucker}\ \emph {et~al.}(2013)\citenamefont
  {H\"ucker}, \citenamefont {v.~Zimmermann}, \citenamefont {Xu}, \citenamefont
  {Wen}, \citenamefont {Gu},\ and\ \citenamefont {Tranquada}}]{Hucker2013}%
  \BibitemOpen
  \bibfield  {author} {\bibinfo {author} {\bibfnamefont {M.}~\bibnamefont
  {H\"ucker}}, \bibinfo {author} {\bibfnamefont {M.}~\bibnamefont
  {v.~Zimmermann}}, \bibinfo {author} {\bibfnamefont {Z.~J.}\ \bibnamefont
  {Xu}}, \bibinfo {author} {\bibfnamefont {J.~S.}\ \bibnamefont {Wen}},
  \bibinfo {author} {\bibfnamefont {G.~D.}\ \bibnamefont {Gu}}, \ and\ \bibinfo
  {author} {\bibfnamefont {J.~M.}\ \bibnamefont {Tranquada}},\ }\href {\doibase
  10.1103/PhysRevB.87.014501} {\bibfield  {journal} {\bibinfo  {journal} {Phys.
  Rev. B}\ }\textbf {\bibinfo {volume} {87}},\ \bibinfo {pages} {014501}
  (\bibinfo {year} {2013})}\BibitemShut {NoStop}%
\bibitem [{\citenamefont {Thampy}\ \emph {et~al.}(2013)\citenamefont {Thampy},
  \citenamefont {Blanco-Canosa}, \citenamefont {Garcia-Fernandez},
  \citenamefont {Dean}, \citenamefont {Gu}, \citenamefont {F\"orst},
  \citenamefont {Loew}, \citenamefont {Keimer}, \citenamefont {Le~Tacon},
  \citenamefont {Wilkins},\ and\ \citenamefont {Hill}}]{Thampy2013}%
  \BibitemOpen
  \bibfield  {author} {\bibinfo {author} {\bibfnamefont {V.}~\bibnamefont
  {Thampy}}, \bibinfo {author} {\bibfnamefont {S.}~\bibnamefont
  {Blanco-Canosa}}, \bibinfo {author} {\bibfnamefont {M.}~\bibnamefont
  {Garcia-Fernandez}}, \bibinfo {author} {\bibfnamefont {M.~P.~M.}\
  \bibnamefont {Dean}}, \bibinfo {author} {\bibfnamefont {G.~D.}\ \bibnamefont
  {Gu}}, \bibinfo {author} {\bibfnamefont {M.}~\bibnamefont {F\"orst}},
  \bibinfo {author} {\bibfnamefont {T.}~\bibnamefont {Loew}}, \bibinfo {author}
  {\bibfnamefont {B.}~\bibnamefont {Keimer}}, \bibinfo {author} {\bibfnamefont
  {M.}~\bibnamefont {Le~Tacon}}, \bibinfo {author} {\bibfnamefont {S.~B.}\
  \bibnamefont {Wilkins}}, \ and\ \bibinfo {author} {\bibfnamefont {J.~P.}\
  \bibnamefont {Hill}},\ }\href {\doibase 10.1103/PhysRevB.88.024505}
  {\bibfield  {journal} {\bibinfo  {journal} {Phys. Rev. B}\ }\textbf {\bibinfo
  {volume} {88}},\ \bibinfo {pages} {024505} (\bibinfo {year}
  {2013})}\BibitemShut {NoStop}%
\bibitem [{\citenamefont {Dean}\ \emph {et~al.}(2013)\citenamefont {Dean},
  \citenamefont {Dellea}, \citenamefont {Minola}, \citenamefont {Wilkins},
  \citenamefont {Konik}, \citenamefont {Gu}, \citenamefont {Le~Tacon},
  \citenamefont {Brookes}, \citenamefont {Yakhou-Harris}, \citenamefont
  {Kummer}, \citenamefont {Hill}, \citenamefont {Braicovich},\ and\
  \citenamefont {Ghiringhelli}}]{DeanLBCO2013}%
  \BibitemOpen
  \bibfield  {author} {\bibinfo {author} {\bibfnamefont {M.~P.~M.}\
  \bibnamefont {Dean}}, \bibinfo {author} {\bibfnamefont {G.}~\bibnamefont
  {Dellea}}, \bibinfo {author} {\bibfnamefont {M.}~\bibnamefont {Minola}},
  \bibinfo {author} {\bibfnamefont {S.~B.}\ \bibnamefont {Wilkins}}, \bibinfo
  {author} {\bibfnamefont {R.~M.}\ \bibnamefont {Konik}}, \bibinfo {author}
  {\bibfnamefont {G.~D.}\ \bibnamefont {Gu}}, \bibinfo {author} {\bibfnamefont
  {M.}~\bibnamefont {Le~Tacon}}, \bibinfo {author} {\bibfnamefont {N.~B.}\
  \bibnamefont {Brookes}}, \bibinfo {author} {\bibfnamefont {F.}~\bibnamefont
  {Yakhou-Harris}}, \bibinfo {author} {\bibfnamefont {K.}~\bibnamefont
  {Kummer}}, \bibinfo {author} {\bibfnamefont {J.~P.}\ \bibnamefont {Hill}},
  \bibinfo {author} {\bibfnamefont {L.}~\bibnamefont {Braicovich}}, \ and\
  \bibinfo {author} {\bibfnamefont {G.}~\bibnamefont {Ghiringhelli}},\ }\href
  {\doibase 10.1103/PhysRevB.88.020403} {\bibfield  {journal} {\bibinfo
  {journal} {Phys. Rev. B}\ }\textbf {\bibinfo {volume} {88}},\ \bibinfo
  {pages} {020403} (\bibinfo {year} {2013})}\BibitemShut {NoStop}%
\bibitem [{\citenamefont {Dainty}(1984)}]{DaintyBook}%
  \BibitemOpen
  \bibfield  {author} {\bibinfo {author} {\bibfnamefont {J.}~\bibnamefont
  {Dainty}},\ }\enquote {\bibinfo {title} {Laser speckle and related
  phenomena},}\ \ (\bibinfo  {publisher} {Springer-Verlag},\ \bibinfo {address}
  {New York},\ \bibinfo {year} {1984})\BibitemShut {NoStop}%
\bibitem [{\citenamefont {Brauer}\ \emph {et~al.}(1995)\citenamefont {Brauer},
  \citenamefont {Stephenson}, \citenamefont {Sutton}, \citenamefont
  {Br\"uning}, \citenamefont {Dufresne}, \citenamefont {Mochrie}, \citenamefont
  {Gr\"ubel}, \citenamefont {Als-Nielsen},\ and\ \citenamefont
  {Abernathy}}]{Brauer1995}%
  \BibitemOpen
  \bibfield  {author} {\bibinfo {author} {\bibfnamefont {S.}~\bibnamefont
  {Brauer}}, \bibinfo {author} {\bibfnamefont {G.~B.}\ \bibnamefont
  {Stephenson}}, \bibinfo {author} {\bibfnamefont {M.}~\bibnamefont {Sutton}},
  \bibinfo {author} {\bibfnamefont {R.}~\bibnamefont {Br\"uning}}, \bibinfo
  {author} {\bibfnamefont {E.}~\bibnamefont {Dufresne}}, \bibinfo {author}
  {\bibfnamefont {S.~G.~J.}\ \bibnamefont {Mochrie}}, \bibinfo {author}
  {\bibfnamefont {G.}~\bibnamefont {Gr\"ubel}}, \bibinfo {author}
  {\bibfnamefont {J.}~\bibnamefont {Als-Nielsen}}, \ and\ \bibinfo {author}
  {\bibfnamefont {D.~L.}\ \bibnamefont {Abernathy}},\ }\href {\doibase
  10.1103/PhysRevLett.74.2010} {\bibfield  {journal} {\bibinfo  {journal}
  {Phys. Rev. Lett.}\ }\textbf {\bibinfo {volume} {74}},\ \bibinfo {pages}
  {2010} (\bibinfo {year} {1995})}\BibitemShut {NoStop}%
\bibitem [{\citenamefont {Sutton}(2008)}]{Sutton2008}%
  \BibitemOpen
  \bibfield  {author} {\bibinfo {author} {\bibfnamefont {M.}~\bibnamefont
  {Sutton}},\ }\href {\doibase http://dx.doi.org/10.1016/j.crhy.2007.04.008}
  {\bibfield  {journal} {\bibinfo  {journal} {Comptes Rendus Physique}\
  }\textbf {\bibinfo {volume} {9}},\ \bibinfo {pages} {657 } (\bibinfo {year}
  {2008})}\BibitemShut {NoStop}%
\bibitem [{\citenamefont {Shpyrko}\ \emph {et~al.}(2007)\citenamefont
  {Shpyrko}, \citenamefont {Isaacs}, \citenamefont {Logan}, \citenamefont
  {Feng}, \citenamefont {Aeppli}, \citenamefont {Jaramillo}, \citenamefont
  {Kim}, \citenamefont {Rosenbaum}, \citenamefont {Zschack}, \citenamefont
  {Sprung} \emph {et~al.}}]{Shpyrko2007}%
  \BibitemOpen
  \bibfield  {author} {\bibinfo {author} {\bibfnamefont {O.}~\bibnamefont
  {Shpyrko}}, \bibinfo {author} {\bibfnamefont {E.}~\bibnamefont {Isaacs}},
  \bibinfo {author} {\bibfnamefont {J.}~\bibnamefont {Logan}}, \bibinfo
  {author} {\bibfnamefont {Y.}~\bibnamefont {Feng}}, \bibinfo {author}
  {\bibfnamefont {G.}~\bibnamefont {Aeppli}}, \bibinfo {author} {\bibfnamefont
  {R.}~\bibnamefont {Jaramillo}}, \bibinfo {author} {\bibfnamefont
  {H.}~\bibnamefont {Kim}}, \bibinfo {author} {\bibfnamefont {T.}~\bibnamefont
  {Rosenbaum}}, \bibinfo {author} {\bibfnamefont {P.}~\bibnamefont {Zschack}},
  \bibinfo {author} {\bibfnamefont {M.}~\bibnamefont {Sprung}},  \emph
  {et~al.},\ }\href@noop {} {\bibfield  {journal} {\bibinfo  {journal}
  {Nature}\ }\textbf {\bibinfo {volume} {447}},\ \bibinfo {pages} {68}
  (\bibinfo {year} {2007})}\BibitemShut {NoStop}%
\bibitem [{\citenamefont {Su}\ \emph {et~al.}(2012)\citenamefont {Su},
  \citenamefont {Sandy}, \citenamefont {Mohanty}, \citenamefont {Shpyrko},\
  and\ \citenamefont {Sutton}}]{Su2012}%
  \BibitemOpen
  \bibfield  {author} {\bibinfo {author} {\bibfnamefont {J.-D.}\ \bibnamefont
  {Su}}, \bibinfo {author} {\bibfnamefont {A.~R.}\ \bibnamefont {Sandy}},
  \bibinfo {author} {\bibfnamefont {J.}~\bibnamefont {Mohanty}}, \bibinfo
  {author} {\bibfnamefont {O.~G.}\ \bibnamefont {Shpyrko}}, \ and\ \bibinfo
  {author} {\bibfnamefont {M.}~\bibnamefont {Sutton}},\ }\href@noop {}
  {\bibfield  {journal} {\bibinfo  {journal} {Physical Review B}\ }\textbf
  {\bibinfo {volume} {86}},\ \bibinfo {pages} {205105} (\bibinfo {year}
  {2012})}\BibitemShut {NoStop}%
\bibitem [{\citenamefont {{M. Sutton}}\ \emph {et~al.}(2002)\citenamefont {{M.
  Sutton}}, \citenamefont {{Y. Li}}, \citenamefont {{J.D. Brock}},\ and\
  \citenamefont {{R.E. Thorne}}}]{sutt02}%
  \BibitemOpen
  \bibfield  {author} {\bibinfo {author} {\bibnamefont {{M. Sutton}}}, \bibinfo
  {author} {\bibnamefont {{Y. Li}}}, \bibinfo {author} {\bibnamefont {{J.D.
  Brock}}}, \ and\ \bibinfo {author} {\bibnamefont {{R.E. Thorne}}},\ }\href
  {\doibase 10.1051/jp4:20020342} {\bibfield  {journal} {\bibinfo  {journal}
  {J. Phys. IV France}\ }\textbf {\bibinfo {volume} {12}},\ \bibinfo {pages}
  {3} (\bibinfo {year} {2002})}\BibitemShut {NoStop}%
\bibitem [{\citenamefont {Pinsolle}\ \emph {et~al.}(2012)\citenamefont
  {Pinsolle}, \citenamefont {Kirova}, \citenamefont {Jacques}, \citenamefont
  {Sinchenko},\ and\ \citenamefont {Le~Bolloc'h}}]{Pinsolle2012}%
  \BibitemOpen
  \bibfield  {author} {\bibinfo {author} {\bibfnamefont {E.}~\bibnamefont
  {Pinsolle}}, \bibinfo {author} {\bibfnamefont {N.}~\bibnamefont {Kirova}},
  \bibinfo {author} {\bibfnamefont {V.~L.~R.}\ \bibnamefont {Jacques}},
  \bibinfo {author} {\bibfnamefont {A.~A.}\ \bibnamefont {Sinchenko}}, \ and\
  \bibinfo {author} {\bibfnamefont {D.}~\bibnamefont {Le~Bolloc'h}},\ }\href
  {\doibase 10.1103/PhysRevLett.109.256402} {\bibfield  {journal} {\bibinfo
  {journal} {Phys. Rev. Lett.}\ }\textbf {\bibinfo {volume} {109}},\ \bibinfo
  {pages} {256402} (\bibinfo {year} {2012})}\BibitemShut {NoStop}%
\bibitem [{\citenamefont {Doering}\ \emph {et~al.}(2011)\citenamefont
  {Doering}, \citenamefont {Chuang}, \citenamefont {Andresen}, \citenamefont
  {Chow}, \citenamefont {Contarato}, \citenamefont {Cummings}, \citenamefont
  {Domning}, \citenamefont {Joseph}, \citenamefont {Pepper}, \citenamefont
  {Smith}, \citenamefont {Zizka}, \citenamefont {Ford}, \citenamefont {Lee},
  \citenamefont {Weaver}, \citenamefont {Patthey}, \citenamefont {Weizeorick},
  \citenamefont {Hussain},\ and\ \citenamefont {Denes}}]{fccd_camera}%
  \BibitemOpen
  \bibfield  {author} {\bibinfo {author} {\bibfnamefont {D.}~\bibnamefont
  {Doering}}, \bibinfo {author} {\bibfnamefont {Y.-D.}\ \bibnamefont {Chuang}},
  \bibinfo {author} {\bibfnamefont {N.}~\bibnamefont {Andresen}}, \bibinfo
  {author} {\bibfnamefont {K.}~\bibnamefont {Chow}}, \bibinfo {author}
  {\bibfnamefont {D.}~\bibnamefont {Contarato}}, \bibinfo {author}
  {\bibfnamefont {C.}~\bibnamefont {Cummings}}, \bibinfo {author}
  {\bibfnamefont {E.}~\bibnamefont {Domning}}, \bibinfo {author} {\bibfnamefont
  {J.}~\bibnamefont {Joseph}}, \bibinfo {author} {\bibfnamefont {J.~S.}\
  \bibnamefont {Pepper}}, \bibinfo {author} {\bibfnamefont {B.}~\bibnamefont
  {Smith}}, \bibinfo {author} {\bibfnamefont {G.}~\bibnamefont {Zizka}},
  \bibinfo {author} {\bibfnamefont {C.}~\bibnamefont {Ford}}, \bibinfo {author}
  {\bibfnamefont {W.~S.}\ \bibnamefont {Lee}}, \bibinfo {author} {\bibfnamefont
  {M.}~\bibnamefont {Weaver}}, \bibinfo {author} {\bibfnamefont
  {L.}~\bibnamefont {Patthey}}, \bibinfo {author} {\bibfnamefont
  {J.}~\bibnamefont {Weizeorick}}, \bibinfo {author} {\bibfnamefont
  {Z.}~\bibnamefont {Hussain}}, \ and\ \bibinfo {author} {\bibfnamefont
  {P.}~\bibnamefont {Denes}},\ }\href
  {http://scitation.aip.org/content/aip/journal/rsi/82/7/10.1063/1.3609862}
  {\bibfield  {journal} {\bibinfo  {journal} {Review of Scientific
  Instruments}\ }\textbf {\bibinfo {volume} {82}},\ \bibinfo {eid} {073303}
  (\bibinfo {year} {2011})}\BibitemShut {NoStop}%
\bibitem [{\citenamefont {Dean}(2015)}]{Dean2015}%
  \BibitemOpen
  \bibfield  {author} {\bibinfo {author} {\bibfnamefont {M.~P.~M.}\
  \bibnamefont {Dean}},\ }\href {\doibase
  http://dx.doi.org/10.1016/j.jmmm.2014.03.057} {\bibfield  {journal} {\bibinfo
   {journal} {Journal of Magnetism and Magnetic Materials}\ }\textbf {\bibinfo
  {volume} {376}},\ \bibinfo {pages} {3 } (\bibinfo {year} {2015})}\BibitemShut
  {NoStop}%
\bibitem [{sup()}]{supplementary}%
  \BibitemOpen
  \href@noop {} {}\bibinfo {note} {See Supplemental Material [url], which
  includes Ref. \cite{SuttonChapter}}\BibitemShut {NoStop}%
\bibitem [{\citenamefont {Shpyrko}(2014)}]{Shpyrko2014}%
  \BibitemOpen
  \bibfield  {author} {\bibinfo {author} {\bibfnamefont {O.~G.}\ \bibnamefont
  {Shpyrko}},\ }\href {\doibase 10.1107/S1600577514018232} {\bibfield
  {journal} {\bibinfo  {journal} {Journal of Synchrotron Radiation}\ }\textbf
  {\bibinfo {volume} {21}},\ \bibinfo {pages} {1057} (\bibinfo {year}
  {2014})}\BibitemShut {NoStop}%
\bibitem [{\citenamefont {Goto}\ \emph {et~al.}(1994)\citenamefont {Goto},
  \citenamefont {Kazama}, \citenamefont {Miyagawa},\ and\ \citenamefont
  {Fukase}}]{goto94}%
  \BibitemOpen
  \bibfield  {author} {\bibinfo {author} {\bibfnamefont {T.}~\bibnamefont
  {Goto}}, \bibinfo {author} {\bibfnamefont {S.}~\bibnamefont {Kazama}},
  \bibinfo {author} {\bibfnamefont {K.}~\bibnamefont {Miyagawa}}, \ and\
  \bibinfo {author} {\bibfnamefont {T.}~\bibnamefont {Fukase}},\ }\href@noop {}
  {\bibfield  {journal} {\bibinfo  {journal} {J. Phys. Soc. Japan}\ }\textbf
  {\bibinfo {volume} {63}},\ \bibinfo {pages} {3494} (\bibinfo {year}
  {1994})}\BibitemShut {NoStop}%
\bibitem [{\citenamefont {Hunt}\ \emph
  {et~al.}(2001{\natexlab{a}})\citenamefont {Hunt}, \citenamefont {Singer},
  \citenamefont {Cederstr\"om},\ and\ \citenamefont {Imai}}]{hunt01}%
  \BibitemOpen
  \bibfield  {author} {\bibinfo {author} {\bibfnamefont {A.~W.}\ \bibnamefont
  {Hunt}}, \bibinfo {author} {\bibfnamefont {P.~M.}\ \bibnamefont {Singer}},
  \bibinfo {author} {\bibfnamefont {A.~F.}\ \bibnamefont {Cederstr\"om}}, \
  and\ \bibinfo {author} {\bibfnamefont {T.}~\bibnamefont {Imai}},\ }\href
  {\doibase 10.1103/PhysRevB.64.134525} {\bibfield  {journal} {\bibinfo
  {journal} {Phys. Rev. B}\ }\textbf {\bibinfo {volume} {64}},\ \bibinfo
  {pages} {134525} (\bibinfo {year} {2001}{\natexlab{a}})}\BibitemShut
  {NoStop}%
\bibitem [{\citenamefont {Julien}\ \emph {et~al.}(2001)\citenamefont {Julien},
  \citenamefont {Campana}, \citenamefont {Rigamonti}, \citenamefont {Carretta},
  \citenamefont {Borsa}, \citenamefont {Kuhns}, \citenamefont {Reyes},
  \citenamefont {Moulton}, \citenamefont {Horvati\ifmmode~\acute{c}\else
  \'{c}\fi{}}, \citenamefont {Berthier}, \citenamefont {Vietkin},\ and\
  \citenamefont {Revcolevschi}}]{juli01}%
  \BibitemOpen
  \bibfield  {author} {\bibinfo {author} {\bibfnamefont {M.-H.}\ \bibnamefont
  {Julien}}, \bibinfo {author} {\bibfnamefont {A.}~\bibnamefont {Campana}},
  \bibinfo {author} {\bibfnamefont {A.}~\bibnamefont {Rigamonti}}, \bibinfo
  {author} {\bibfnamefont {P.}~\bibnamefont {Carretta}}, \bibinfo {author}
  {\bibfnamefont {F.}~\bibnamefont {Borsa}}, \bibinfo {author} {\bibfnamefont
  {P.}~\bibnamefont {Kuhns}}, \bibinfo {author} {\bibfnamefont {A.~P.}\
  \bibnamefont {Reyes}}, \bibinfo {author} {\bibfnamefont {W.~G.}\ \bibnamefont
  {Moulton}}, \bibinfo {author} {\bibfnamefont {M.}~\bibnamefont
  {Horvati\ifmmode~\acute{c}\else \'{c}\fi{}}}, \bibinfo {author}
  {\bibfnamefont {C.}~\bibnamefont {Berthier}}, \bibinfo {author}
  {\bibfnamefont {A.}~\bibnamefont {Vietkin}}, \ and\ \bibinfo {author}
  {\bibfnamefont {A.}~\bibnamefont {Revcolevschi}},\ }\href {\doibase
  10.1103/PhysRevB.63.144508} {\bibfield  {journal} {\bibinfo  {journal} {Phys.
  Rev. B}\ }\textbf {\bibinfo {volume} {63}},\ \bibinfo {pages} {144508}
  (\bibinfo {year} {2001})}\BibitemShut {NoStop}%
\bibitem [{\citenamefont {{Pelc}}\ \emph {et~al.}(2016)\citenamefont {{Pelc}},
  \citenamefont {{Grafe}}, \citenamefont {{Gu}},\ and\ \citenamefont {{Po{\v
  z}ek}}}]{Pelc2016}%
  \BibitemOpen
  \bibfield  {author} {\bibinfo {author} {\bibfnamefont {D.}~\bibnamefont
  {{Pelc}}}, \bibinfo {author} {\bibfnamefont {H.-J.}\ \bibnamefont {{Grafe}}},
  \bibinfo {author} {\bibfnamefont {G.}~\bibnamefont {{Gu}}}, \ and\ \bibinfo
  {author} {\bibfnamefont {M.}~\bibnamefont {{Po{\v z}ek}}},\ }\href@noop {}
  {\bibfield  {journal} {\bibinfo  {journal} {ArXiv e-prints}\ } (\bibinfo
  {year} {2016})},\ \Eprint {http://arxiv.org/abs/1609.02162}
  {arXiv:1609.02162} \BibitemShut {NoStop}%
\bibitem [{\citenamefont {Parker}\ \emph {et~al.}(2010)\citenamefont {Parker},
  \citenamefont {Aynajian}, \citenamefont {da~Silva~Neto}, \citenamefont
  {Pushp}, \citenamefont {Ono}, \citenamefont {Wen}, \citenamefont {Xu},
  \citenamefont {Gu},\ and\ \citenamefont {Yazdani}}]{Parker2010}%
  \BibitemOpen
  \bibfield  {author} {\bibinfo {author} {\bibfnamefont {C.~V.}\ \bibnamefont
  {Parker}}, \bibinfo {author} {\bibfnamefont {P.}~\bibnamefont {Aynajian}},
  \bibinfo {author} {\bibfnamefont {E.~H.}\ \bibnamefont {da~Silva~Neto}},
  \bibinfo {author} {\bibfnamefont {A.}~\bibnamefont {Pushp}}, \bibinfo
  {author} {\bibfnamefont {S.}~\bibnamefont {Ono}}, \bibinfo {author}
  {\bibfnamefont {J.}~\bibnamefont {Wen}}, \bibinfo {author} {\bibfnamefont
  {Z.}~\bibnamefont {Xu}}, \bibinfo {author} {\bibfnamefont {G.}~\bibnamefont
  {Gu}}, \ and\ \bibinfo {author} {\bibfnamefont {A.}~\bibnamefont {Yazdani}},\
  }\href {http://dx.doi.org/10.1038/nature09597} {\bibfield  {journal}
  {\bibinfo  {journal} {Nature}\ }\textbf {\bibinfo {volume} {468}},\ \bibinfo
  {pages} {677} (\bibinfo {year} {2010})}\BibitemShut {NoStop}%
\bibitem [{\citenamefont {da~Silva~Neto}\ \emph {et~al.}(2015)\citenamefont
  {da~Silva~Neto}, \citenamefont {Comin}, \citenamefont {He}, \citenamefont
  {Sutarto}, \citenamefont {Jiang}, \citenamefont {Greene}, \citenamefont
  {Sawatzky},\ and\ \citenamefont {Damascelli}}]{daSilvaNeto2015}%
  \BibitemOpen
  \bibfield  {author} {\bibinfo {author} {\bibfnamefont {E.~H.}\ \bibnamefont
  {da~Silva~Neto}}, \bibinfo {author} {\bibfnamefont {R.}~\bibnamefont
  {Comin}}, \bibinfo {author} {\bibfnamefont {F.}~\bibnamefont {He}}, \bibinfo
  {author} {\bibfnamefont {R.}~\bibnamefont {Sutarto}}, \bibinfo {author}
  {\bibfnamefont {Y.}~\bibnamefont {Jiang}}, \bibinfo {author} {\bibfnamefont
  {R.~L.}\ \bibnamefont {Greene}}, \bibinfo {author} {\bibfnamefont {G.~A.}\
  \bibnamefont {Sawatzky}}, \ and\ \bibinfo {author} {\bibfnamefont
  {A.}~\bibnamefont {Damascelli}},\ }\href@noop {} {\bibfield  {journal}
  {\bibinfo  {journal} {Science}\ }\textbf {\bibinfo {volume} {347}},\ \bibinfo
  {pages} {282} (\bibinfo {year} {2015})}\BibitemShut {NoStop}%
\bibitem [{\citenamefont {Valla}\ \emph {et~al.}(2006)\citenamefont {Valla},
  \citenamefont {Fedorov}, \citenamefont {Lee}, \citenamefont {Davis},\ and\
  \citenamefont {Gu}}]{Valla1914}%
  \BibitemOpen
  \bibfield  {author} {\bibinfo {author} {\bibfnamefont {T.}~\bibnamefont
  {Valla}}, \bibinfo {author} {\bibfnamefont {A.~V.}\ \bibnamefont {Fedorov}},
  \bibinfo {author} {\bibfnamefont {J.}~\bibnamefont {Lee}}, \bibinfo {author}
  {\bibfnamefont {J.~C.}\ \bibnamefont {Davis}}, \ and\ \bibinfo {author}
  {\bibfnamefont {G.~D.}\ \bibnamefont {Gu}},\ }\href {\doibase
  10.1126/science.1134742} {\bibfield  {journal} {\bibinfo  {journal}
  {Science}\ }\textbf {\bibinfo {volume} {314}},\ \bibinfo {pages} {1914}
  (\bibinfo {year} {2006})}\BibitemShut {NoStop}%
\bibitem [{\citenamefont {Achkar}\ \emph
  {et~al.}(2016{\natexlab{a}})\citenamefont {Achkar}, \citenamefont {Zwiebler},
  \citenamefont {McMahon}, \citenamefont {He}, \citenamefont {Sutarto},
  \citenamefont {Djianto}, \citenamefont {Hao}, \citenamefont {Gingras},
  \citenamefont {H{\"u}cker}, \citenamefont {Gu} \emph {et~al.}}]{Achkar2016}%
  \BibitemOpen
  \bibfield  {author} {\bibinfo {author} {\bibfnamefont {A.}~\bibnamefont
  {Achkar}}, \bibinfo {author} {\bibfnamefont {M.}~\bibnamefont {Zwiebler}},
  \bibinfo {author} {\bibfnamefont {C.}~\bibnamefont {McMahon}}, \bibinfo
  {author} {\bibfnamefont {F.}~\bibnamefont {He}}, \bibinfo {author}
  {\bibfnamefont {R.}~\bibnamefont {Sutarto}}, \bibinfo {author} {\bibfnamefont
  {I.}~\bibnamefont {Djianto}}, \bibinfo {author} {\bibfnamefont
  {Z.}~\bibnamefont {Hao}}, \bibinfo {author} {\bibfnamefont {M.~J.}\
  \bibnamefont {Gingras}}, \bibinfo {author} {\bibfnamefont {M.}~\bibnamefont
  {H{\"u}cker}}, \bibinfo {author} {\bibfnamefont {G.}~\bibnamefont {Gu}},
  \emph {et~al.},\ }\href@noop {} {\bibfield  {journal} {\bibinfo  {journal}
  {Science}\ }\textbf {\bibinfo {volume} {351}},\ \bibinfo {pages} {576}
  (\bibinfo {year} {2016}{\natexlab{a}})}\BibitemShut {NoStop}%
\bibitem [{\citenamefont {Tranquada}\ \emph {et~al.}(2008)\citenamefont
  {Tranquada}, \citenamefont {Gu}, \citenamefont {H\"ucker}, \citenamefont
  {Jie}, \citenamefont {Kang}, \citenamefont {Klingeler}, \citenamefont {Li},
  \citenamefont {Tristan}, \citenamefont {Wen}, \citenamefont {Xu},
  \citenamefont {Xu}, \citenamefont {Zhou},\ and\ \citenamefont
  {v.~Zimmermann}}]{Tranquada2008}%
  \BibitemOpen
  \bibfield  {author} {\bibinfo {author} {\bibfnamefont {J.~M.}\ \bibnamefont
  {Tranquada}}, \bibinfo {author} {\bibfnamefont {G.~D.}\ \bibnamefont {Gu}},
  \bibinfo {author} {\bibfnamefont {M.}~\bibnamefont {H\"ucker}}, \bibinfo
  {author} {\bibfnamefont {Q.}~\bibnamefont {Jie}}, \bibinfo {author}
  {\bibfnamefont {H.-J.}\ \bibnamefont {Kang}}, \bibinfo {author}
  {\bibfnamefont {R.}~\bibnamefont {Klingeler}}, \bibinfo {author}
  {\bibfnamefont {Q.}~\bibnamefont {Li}}, \bibinfo {author} {\bibfnamefont
  {N.}~\bibnamefont {Tristan}}, \bibinfo {author} {\bibfnamefont {J.~S.}\
  \bibnamefont {Wen}}, \bibinfo {author} {\bibfnamefont {G.~Y.}\ \bibnamefont
  {Xu}}, \bibinfo {author} {\bibfnamefont {Z.~J.}\ \bibnamefont {Xu}}, \bibinfo
  {author} {\bibfnamefont {J.}~\bibnamefont {Zhou}}, \ and\ \bibinfo {author}
  {\bibfnamefont {M.}~\bibnamefont {v.~Zimmermann}},\ }\href {\doibase
  10.1103/PhysRevB.78.174529} {\bibfield  {journal} {\bibinfo  {journal} {Phys.
  Rev. B}\ }\textbf {\bibinfo {volume} {78}},\ \bibinfo {pages} {174529}
  (\bibinfo {year} {2008})}\BibitemShut {NoStop}%
\bibitem [{\citenamefont {Savici}\ \emph {et~al.}(2005)\citenamefont {Savici},
  \citenamefont {Fukaya}, \citenamefont {Gat-Malureanu}, \citenamefont {Ito},
  \citenamefont {Russo}, \citenamefont {Uemura}, \citenamefont {Wiebe},
  \citenamefont {Kyriakou}, \citenamefont {MacDougall}, \citenamefont {Rovers},
  \citenamefont {Luke}, \citenamefont {Kojima}, \citenamefont {Goto},
  \citenamefont {Uchida}, \citenamefont {Kadono}, \citenamefont {Yamada},
  \citenamefont {Tajima}, \citenamefont {Masui}, \citenamefont {Eisaki},
  \citenamefont {Kaneko}, \citenamefont {Greven},\ and\ \citenamefont
  {Gu}}]{savi05}%
  \BibitemOpen
  \bibfield  {author} {\bibinfo {author} {\bibfnamefont {A.~T.}\ \bibnamefont
  {Savici}}, \bibinfo {author} {\bibfnamefont {A.}~\bibnamefont {Fukaya}},
  \bibinfo {author} {\bibfnamefont {I.~M.}\ \bibnamefont {Gat-Malureanu}},
  \bibinfo {author} {\bibfnamefont {T.}~\bibnamefont {Ito}}, \bibinfo {author}
  {\bibfnamefont {P.~L.}\ \bibnamefont {Russo}}, \bibinfo {author}
  {\bibfnamefont {Y.~J.}\ \bibnamefont {Uemura}}, \bibinfo {author}
  {\bibfnamefont {C.~R.}\ \bibnamefont {Wiebe}}, \bibinfo {author}
  {\bibfnamefont {P.~P.}\ \bibnamefont {Kyriakou}}, \bibinfo {author}
  {\bibfnamefont {G.~J.}\ \bibnamefont {MacDougall}}, \bibinfo {author}
  {\bibfnamefont {M.~T.}\ \bibnamefont {Rovers}}, \bibinfo {author}
  {\bibfnamefont {G.~M.}\ \bibnamefont {Luke}}, \bibinfo {author}
  {\bibfnamefont {K.~M.}\ \bibnamefont {Kojima}}, \bibinfo {author}
  {\bibfnamefont {M.}~\bibnamefont {Goto}}, \bibinfo {author} {\bibfnamefont
  {S.}~\bibnamefont {Uchida}}, \bibinfo {author} {\bibfnamefont
  {R.}~\bibnamefont {Kadono}}, \bibinfo {author} {\bibfnamefont
  {K.}~\bibnamefont {Yamada}}, \bibinfo {author} {\bibfnamefont
  {S.}~\bibnamefont {Tajima}}, \bibinfo {author} {\bibfnamefont
  {T.}~\bibnamefont {Masui}}, \bibinfo {author} {\bibfnamefont
  {H.}~\bibnamefont {Eisaki}}, \bibinfo {author} {\bibfnamefont
  {N.}~\bibnamefont {Kaneko}}, \bibinfo {author} {\bibfnamefont
  {M.}~\bibnamefont {Greven}}, \ and\ \bibinfo {author} {\bibfnamefont {G.~D.}\
  \bibnamefont {Gu}},\ }\href@noop {} {\bibfield  {journal} {\bibinfo
  {journal} {Phys. Rev. Lett.}\ }\textbf {\bibinfo {volume} {95}},\ \bibinfo
  {eid} {157001} (\bibinfo {year} {2005})}\BibitemShut {NoStop}%
\bibitem [{\citenamefont {Hunt}\ \emph
  {et~al.}(2001{\natexlab{b}})\citenamefont {Hunt}, \citenamefont {Singer},
  \citenamefont {Cederstr\"om},\ and\ \citenamefont {Imai}}]{Hunt2001}%
  \BibitemOpen
  \bibfield  {author} {\bibinfo {author} {\bibfnamefont {A.~W.}\ \bibnamefont
  {Hunt}}, \bibinfo {author} {\bibfnamefont {P.~M.}\ \bibnamefont {Singer}},
  \bibinfo {author} {\bibfnamefont {A.~F.}\ \bibnamefont {Cederstr\"om}}, \
  and\ \bibinfo {author} {\bibfnamefont {T.}~\bibnamefont {Imai}},\ }\href
  {\doibase 10.1103/PhysRevB.64.134525} {\bibfield  {journal} {\bibinfo
  {journal} {Phys. Rev. B}\ }\textbf {\bibinfo {volume} {64}},\ \bibinfo
  {pages} {134525} (\bibinfo {year} {2001}{\natexlab{b}})}\BibitemShut
  {NoStop}%
\bibitem [{\citenamefont {Achkar}\ \emph
  {et~al.}(2016{\natexlab{b}})\citenamefont {Achkar}, \citenamefont {He},
  \citenamefont {Sutarto}, \citenamefont {McMahon}, \citenamefont {Zwiebler},
  \citenamefont {Hucker}, \citenamefont {Gu}, \citenamefont {Liang},
  \citenamefont {Bonn}, \citenamefont {Hardy}, \citenamefont {Geck},\ and\
  \citenamefont {Hawthorn}}]{achk16b}%
  \BibitemOpen
  \bibfield  {author} {\bibinfo {author} {\bibfnamefont {A.~J.}\ \bibnamefont
  {Achkar}}, \bibinfo {author} {\bibfnamefont {F.}~\bibnamefont {He}}, \bibinfo
  {author} {\bibfnamefont {R.}~\bibnamefont {Sutarto}}, \bibinfo {author}
  {\bibfnamefont {C.}~\bibnamefont {McMahon}}, \bibinfo {author} {\bibfnamefont
  {M.}~\bibnamefont {Zwiebler}}, \bibinfo {author} {\bibfnamefont
  {M.}~\bibnamefont {Hucker}}, \bibinfo {author} {\bibfnamefont {G.~D.}\
  \bibnamefont {Gu}}, \bibinfo {author} {\bibfnamefont {R.}~\bibnamefont
  {Liang}}, \bibinfo {author} {\bibfnamefont {D.~A.}\ \bibnamefont {Bonn}},
  \bibinfo {author} {\bibfnamefont {W.~N.}\ \bibnamefont {Hardy}}, \bibinfo
  {author} {\bibfnamefont {J.}~\bibnamefont {Geck}}, \ and\ \bibinfo {author}
  {\bibfnamefont {D.~G.}\ \bibnamefont {Hawthorn}},\ }\href
  {http://dx.doi.org/10.1038/nmat4568} {\bibfield  {journal} {\bibinfo
  {journal} {Nat. Mater.}\ }\textbf {\bibinfo {volume} {15}},\ \bibinfo {pages}
  {616} (\bibinfo {year} {2016}{\natexlab{b}})}\BibitemShut {NoStop}%
\bibitem [{\citenamefont {H\"ucker}\ \emph {et~al.}(2014)\citenamefont
  {H\"ucker}, \citenamefont {Christensen}, \citenamefont {Holmes},
  \citenamefont {Blackburn}, \citenamefont {Forgan}, \citenamefont {Liang},
  \citenamefont {Bonn}, \citenamefont {Hardy}, \citenamefont {Gutowski},
  \citenamefont {Zimmermann}, \citenamefont {Hayden},\ and\ \citenamefont
  {Chang}}]{huck14}%
  \BibitemOpen
  \bibfield  {author} {\bibinfo {author} {\bibfnamefont {M.}~\bibnamefont
  {H\"ucker}}, \bibinfo {author} {\bibfnamefont {N.~B.}\ \bibnamefont
  {Christensen}}, \bibinfo {author} {\bibfnamefont {A.~T.}\ \bibnamefont
  {Holmes}}, \bibinfo {author} {\bibfnamefont {E.}~\bibnamefont {Blackburn}},
  \bibinfo {author} {\bibfnamefont {E.~M.}\ \bibnamefont {Forgan}}, \bibinfo
  {author} {\bibfnamefont {R.}~\bibnamefont {Liang}}, \bibinfo {author}
  {\bibfnamefont {D.~A.}\ \bibnamefont {Bonn}}, \bibinfo {author}
  {\bibfnamefont {W.~N.}\ \bibnamefont {Hardy}}, \bibinfo {author}
  {\bibfnamefont {O.}~\bibnamefont {Gutowski}}, \bibinfo {author}
  {\bibfnamefont {M.~v.}\ \bibnamefont {Zimmermann}}, \bibinfo {author}
  {\bibfnamefont {S.~M.}\ \bibnamefont {Hayden}}, \ and\ \bibinfo {author}
  {\bibfnamefont {J.}~\bibnamefont {Chang}},\ }\href@noop {} {\bibfield
  {journal} {\bibinfo  {journal} {Phys. Rev. B}\ }\textbf {\bibinfo {volume}
  {90}},\ \bibinfo {pages} {054514} (\bibinfo {year} {2014})}\BibitemShut
  {NoStop}%
\bibitem [{\citenamefont {Jacobsen}\ \emph {et~al.}(2015)\citenamefont
  {Jacobsen}, \citenamefont {Zaliznyak}, \citenamefont {Savici}, \citenamefont
  {Winn}, \citenamefont {Chang}, \citenamefont {H\"ucker}, \citenamefont {Gu},\
  and\ \citenamefont {Tranquada}}]{jaco15}%
  \BibitemOpen
  \bibfield  {author} {\bibinfo {author} {\bibfnamefont {H.}~\bibnamefont
  {Jacobsen}}, \bibinfo {author} {\bibfnamefont {I.~A.}\ \bibnamefont
  {Zaliznyak}}, \bibinfo {author} {\bibfnamefont {A.~T.}\ \bibnamefont
  {Savici}}, \bibinfo {author} {\bibfnamefont {B.~L.}\ \bibnamefont {Winn}},
  \bibinfo {author} {\bibfnamefont {S.}~\bibnamefont {Chang}}, \bibinfo
  {author} {\bibfnamefont {M.}~\bibnamefont {H\"ucker}}, \bibinfo {author}
  {\bibfnamefont {G.~D.}\ \bibnamefont {Gu}}, \ and\ \bibinfo {author}
  {\bibfnamefont {J.~M.}\ \bibnamefont {Tranquada}},\ }\href {\doibase
  10.1103/PhysRevB.92.174525} {\bibfield  {journal} {\bibinfo  {journal} {Phys.
  Rev. B}\ }\textbf {\bibinfo {volume} {92}},\ \bibinfo {pages} {174525}
  (\bibinfo {year} {2015})}\BibitemShut {NoStop}%
\bibitem [{\citenamefont {Hamidian}\ \emph {et~al.}(2016)\citenamefont
  {Hamidian}, \citenamefont {Edkins}, \citenamefont {Joo}, \citenamefont
  {Kostin}, \citenamefont {Eisaki}, \citenamefont {Uchida}, \citenamefont
  {Lawler}, \citenamefont {Kim}, \citenamefont {Mackenzie}, \citenamefont
  {Fujita} \emph {et~al.}}]{Hamidian2016}%
  \BibitemOpen
  \bibfield  {author} {\bibinfo {author} {\bibfnamefont {M.}~\bibnamefont
  {Hamidian}}, \bibinfo {author} {\bibfnamefont {S.}~\bibnamefont {Edkins}},
  \bibinfo {author} {\bibfnamefont {S.~H.}\ \bibnamefont {Joo}}, \bibinfo
  {author} {\bibfnamefont {A.}~\bibnamefont {Kostin}}, \bibinfo {author}
  {\bibfnamefont {H.}~\bibnamefont {Eisaki}}, \bibinfo {author} {\bibfnamefont
  {S.}~\bibnamefont {Uchida}}, \bibinfo {author} {\bibfnamefont
  {M.}~\bibnamefont {Lawler}}, \bibinfo {author} {\bibfnamefont {E.-A.}\
  \bibnamefont {Kim}}, \bibinfo {author} {\bibfnamefont {A.}~\bibnamefont
  {Mackenzie}}, \bibinfo {author} {\bibfnamefont {K.}~\bibnamefont {Fujita}},
  \emph {et~al.},\ }\href@noop {} {\bibfield  {journal} {\bibinfo  {journal}
  {Nature}\ }\textbf {\bibinfo {volume} {532}},\ \bibinfo {pages} {343}
  (\bibinfo {year} {2016})}\BibitemShut {NoStop}%
\bibitem [{\citenamefont {Sutton}(2002)}]{SuttonChapter}%
  \BibitemOpen
  \bibfield  {author} {\bibinfo {author} {\bibfnamefont {M.}~\bibnamefont
  {Sutton}},\ }\enquote {\bibinfo {title} {Coherent x-ray diffraction},}\ in\
  \href@noop {} {\emph {\bibinfo {booktitle} {Third-Generation Hard X-ray
  Synchrotron Radiation Sources: Source Properties, Optics, and Experimental
  Techniques}}},\ \bibinfo {editor} {edited by\ \bibinfo {editor}
  {\bibfnamefont {D.~M.}\ \bibnamefont {Mills}}}\ (\bibinfo  {publisher} {John
  Wiley and Sons},\ \bibinfo {address} {New York},\ \bibinfo {year}
  {2002})\BibitemShut {NoStop}%
\end{thebibliography}
\end{document}